\documentclass[a4paper,11pt]{article}
\pdfoutput=1 % if your are submitting a pdflatex (i.e. if you have
\usepackage{jheppub} % for details on the use of the package, please
\usepackage[T1]{fontenc} % if needed

\usepackage{graphicx}% Include figure files
\usepackage{dcolumn}% Align table columns on decimal point
\usepackage{bm}% bold math
\usepackage{amsmath}
\usepackage{mathrsfs}
\usepackage{bbm}
\usepackage{comment}
\usepackage[shortlabels]{enumitem}
\title{A microscopic analogue of the BMS group}

\author[a]{Daniel Alexander Weiss}

\affiliation[a]{Arnold Sommerfeld Center for Theoretical Physics, Theresienstr. 37, 80333 München, Germany}

\emailAdd{weiss.daniel@physik.uni-muenchen.de}

\abstract{We consider a microscopic analogue of the BMS analysis of asymptotic symmetries by analysing universal geometric structures on infinitesimal tangent light cones. Thereby, two natural microscopic symmetry groups arise: A non-trivially represented Lorentz group and a BMS-like group. The latter has a rich mathematical structure, since it contains the former as a non-canonical subgroup, next to infinitely many other Lorentz subgroups. None of those Lorentz subgroups appears to be intrinsically preferred, and hence, the microscopic BMS-like group constitutes a natural symmetry group for infinitesimal tangent light cones. We compare our investigation with the classical BMS analysis and show, that the microscopic BMS-like group is a gauge group for the bundle of null vectors. Motivated by the various applications of the original BMS group, our findings could have interesting implications: They identify a geometric structure that could be suitable for a bulk analysis of gravitational waves, they suggest a possible enlargement of the fundamental gauge group of gravity and they motivate the possibility of an interrelation between the UV structure of gauge theories, gravitational memory effects and BMS-like symmetries. Also, our results imply, that BMS-like groups arise not only as macroscopic, asymptotic symmetry groups in cosmology, but describe also a fundamental and seemingly unknown microscopic symmetry of pseudo-Riemannian geometry.}

\makeatletter
\gdef\@fpheader{}
\makeatother

\begin{document} 
\maketitle
\section{Introduction}\label{1}

In their seminal works from 1962 (cf. \cite{Bo60, Sa62, Bo62})  Bondi, Metzner, van der Burg and Sachs derived among other results, that the asymptotic symmetry group of asymptotically flat spacetimes at null infinity is not given by the Poincaré group, but by an infinite dimensional generalization, that goes today under the name of the Bondi-Metzner-Sachs (BMS) group. This discovery not only laid the foundation for a coordinate invariant analysis of gravitational waves (cf. \cite{Bo60, Sa62}), but stimulated a lot of research till today: It found application in numerical relativity (cf. \cite{Win12, Bis16}), led more recently to the formulation of celestial holography (cf. e.g. \cite{Pas22}) and even triggered deep insights into the structure of gauge theories on asymptotically flat spacetimes (\cite{Str18}). In a modern language (cf. \cite{Ash14, Ash18}), the original BMS-analysis can be most easily understood by performing a conformal compactification (\cite{Pen65}) of the considered spacetime. In this scenario, null infinity becomes a 3-dimensional null manifold which represents a boundary of the compactified bulk spacetime. As such, it inherits several universal geometric structures from the bulk, that are independent of the specific spacetime under consideration. The symmetry group of those structures is then given by the BMS-group, which describes the asymptotic symmetries that are encompassed by all asymptotically flat spacetimes (cf. \cite{Ash14, Ash18}).

The classic BMS group can hence be understood as a macroscopic symmetry group. It encodes, how spacetimes behave asymptotically at the largest scales. But one could also wonder about the microscopic asymptotics and could hence ask the opposite question: Is there a microscopic analogue of the BMS group which describes, how spacetimes behave asymptotically at microscopic scales? On the first glimpse, this question seems to be trivial. The  Einstein equivalence principle (cf. e.g. \cite{Car19}) tells us directly, that every spacetime behaves microscopically like flat Minkowski space. Hence, the microscopic symmetry group of general relativity should be given by the Lorentz group. This is of course true in some sense and can be formalized in terms of the vielbein formalism (cf. e.g. \cite{Car19}), where the Lorentz group appears as a microscopic gauge freedom in the choice of local inertial frames. But if one compares this with the classic BMS analysis, one will realize, that this is not the correct microscopic analogue the original BMS analysis.

In the original BMS analysis, solely null infinity was analyzed (cf. \cite{Ash14, Ash18}), since gravitational waves are assumed to travel along null rays. Spacelike or timelike directions were discarded. Hence, a microscopic analogue of the classical BMS-analysis should also focus just on microscopic null directions and should consider a microscopic analogue of null infinity. Or more precisely: One should identify the universal structures, that are induced on a natural microscopic null surface that is common to all spacetimes and determine their symmetry group. But which natural microscopic null surface exists in any spacetime? Here again the Einstein equivalence principle comes in, as it states, that any spacetime behaves microscopically (or better, infinitesimally) like Minkowski space. This criterion could hence be understood as a microscopic analogue of asymptotic flatness, since it says, that any spacetime behaves in the microscopic limit asymptotically like flat spacetime. And precisely, as (macroscopic) asymptotic flatness singles out null infinity as the natural macroscopic null surface for the classic BMS analysis, the Einstein equivalence principle, interpreted as a microscopic asymptotic flatness criterion, singles out infinitesimal tangent light cones as natural candidates for a microscopic BMS analysis. Consequently, one can ask in full analogy to the macroscopic case: Which universal structures on infinitesimal tangent light cones are common to all spacetimes satisfying the Einstein equivalence principle, what are the symmetries of those structures and what is the structure and the interpretation of the occuring symmetry groups? Those questions will be answered in this article and thereby we will uncover some interesting structures hidden in the theory of relativity. Especially, we will show as a main result, that a microscopic symmetry group appears, whose structure resembles the original BMS group and which is eligible as a gauge group for the bundle of null vectors for a generic spacetime.  

After this overview of the motivation and philosophy behind our work, we would like to present now its line of argumentation in a concise way. Therefore recall first, that usually the set of null vectors is "identified" with its space of directions, which is commonly denoted as the celestial sphere. Under this identification, the Lorentz group is isomorphic to the Möbius group of conformal automorphisms of the Riemann sphere (cf. e.g. \cite{Pen84, Obl15}). Nevertheless we will see, that null vectors transform under Lorentz transformations not only by a change of their direction, but also by a rescaling of their "length". This behaviour is discarded by the common "identification" of the set of null vectors with the celestial sphere and one could ask, if there is some non-trivial information hidden in those rescalings. At a first glimpse, one might be tempted to think, that no non-trivial coordinate invariant information could be encoded in the length - and consequently also in rescalings - of null vectors, since the metric is degenerate for them. Moreover, the above mentioned isomorphism between the Lorentz and the Möbius group suggests, that there is no interesting information left, which could be encoded in non-trivial rescalings. But this intuition is only partially correct: We will see, that the metric constitutes a kind of distance function on each tangent light cone, although it is degenerate thereon. Metric degeneracy translates then to the statement, that the distance between two null vectors pointing in the same direction is zero. This distance function can then be related to a degenerate Riemannian metric on each tangent light cone and constitutes, together with other, more elementary properties inherited from the ambient tangent space, a set of universal geometric structures existent on any infinitesimal light cone. Those structures are then a microscopic analogue of the macroscopic universal structures at null infinity identified in the original BMS analysis.

One can then start to analyze those automorphisms of tangent light cones, which preserve the identified universal structures and thereby a basic question appears: How should the metric, that appears along above lines on any infinitesimal tangent light cone, be interpreted? As a fixed degenerate metric or as a representative of a conformal equivalence class of degenerate metrics? The former interpretation will lead to a group of isometries, while the latter will give rise to a conformal automorphism group. More or less surprisingly, the non-trivial rescalings of null vectors will appear again if one analyses the isometry group: Those rescalings are needed to compensate the appearing conformal factor in the metric on the light cone, which is induced by the action of Möbius transformations on the Riemann sphere. 

If one analyses then the mathematical structure of the conformal automorphism group, some interesting properties appear: It can be written  as a semidirect product of the Möbius group with a group of smooth, real valued functions on the Riemann sphere, which resembles the structure of the original BMS group. Moreover, it incorporates infinitely many Lorentz subgroups. The isometry group constitutes then just one of those Lorentz subgroups and can be shown to be induced by the standard representation of the Lorentz group on a single tangent space. By investigating those Lorentz subgroups more thoroughly, one realizes, that each Lorentz subgroup singles out a class of length gauges for null vectors, i.e. they set a scale for null vectors. For example, the subgroup of isometries is associated with the $3$-length of null vectors in a vielbein frame. But this seems unsatisfactory: As said before, by metric degeneracy there is no preferred notion of length associated with a single null direction and by this, no structure on the light cone should single out a scale for null vectors. This suggests, that intrinsically no Lorentz subgroup of the conformal automorphism group is preferred and by this it seems, that the conformal automorphism group is intrinsically the more natural symmetry group for tangent light cones. From this point of view, we will then get also a different perspective on the conformal automorphism group: If one tries to introduce meaningful notions of length for null vectors, one necessarily has to enlarge the class of allowed coordinate systems, such that all possible length gauges are included. By doing so, also the symmetry group gets enlarged to the conformal automorphism group, which describes then rescalings of null vector lengths and associated Lorentz transformation laws in an invariant way. But by this, the original Lorentz group action on null vectors looses its distinguished role and infinitely many Lorentz transformation laws appear. This is a way, in which the microscopic BMS-like conformal automorphism group could be interpreted.

Now one could ask, what benefit one could draw from the existence of such a group. Therefore we will show finally, that the isomtery group and the conformal automorphism group are suitable as gauge groups for the bundle of future pointing null vectors. By this we have especially identified a geometric entity, that exists on the bulk of any spacetime and is associated with a BMS-like group. The benefit of this structure has of course to be proven in the future, but motivated by the various applications of the original BMS group we think, that this finding could have interesting implications: On the one hand, the existence of the BMS-like gauge group on the bundle of tangent light cones could lead to a bulk counterpart of the original BMS-anaylsis of gravitational waves and rises in addition questions regarding the fundamental symmetry group of gravity. On the other hand, given the recent discovery of connections between the BMS group, soft theorems and memory effects denoted commonly as the "IR-triangle" \cite{Str16, Str18}, our findings motivate the question, if there could exist an analogous "UV-triangle".

Finally we would like to mention, that the original BMS analysis incorporated of course not only the investigation of universal structures at null infinity and their symmetries, but also the examination of induced higher order structures, dynamical considerations and the analysis of gravitational waves. The microscopic analogues of those questions are not analysed in this article and will be, as sketched in the last paragraph, an object of future research.

\paragraph{Organization of the article:} In section \ref{TL} we will introduce the light cone bundle of a generic spacetime and will identify the universal geometric structures that are induced on infinitesimal tangent light cones by the ambient spacetime geometry as motivated by Einstein's equivalence principle. Thereby we will also explain concisely, how null vectors rescale under standard Lorentz transformations. In section \ref{4} we will analyze those automorphisms of a single infinitesimal tangent light cone, that preserve the universal structure either up to isometry or up to conformal equivalence. In addition, we will define the corresponding automorphism groups. In section \ref{5} we will analyze the structure of those automorphism groups thoroughly. Especially we will show, that the conformal automorphism group can be written as a right semidirect product group that contains infinitely many Lorentz subgroups. We will also explain, how those Lorentz subgroups can be parametrized and how the subgroup of isometries arises as a specific Lorentz subgroup. Moreover we will explain, how Lorentz subgroups correspond to length gauges for null vectors and why this motivates the claim, that the conformal automorphism group could be a more natural symmetry group for tangent light cones than any Lorentz subgroup. Dually we will understand thereby, that the conformal automorphism group could be interpreted as a group which encodes all possible length gauges for null vectors as well as their Lorentz transformation properties. In section \ref{ConcBMS} we will compare our analysis with the original BMS analysis, which will justify, why the conformal automorphism group is called a microscopic analogue of the BMS group. In section \ref{GG} we will sketch, how the conformal automorphism group and its isometry subgroup constitute gauge groups for the light cone bundle. In section \ref{Conc} we will conclude the article by summarizing its findings and by discussing possible implications as well as remaining open questions. In appendix \ref{LMCA} we will review some basic facts on the Riemann sphere and its automorphisms. Moreover we will derive a convenient representation of coordinate systems of the light cone and will derive a transformation law (including rescalings) for null vectors under Lorentz transformations. In appendix \ref{BGT} we will review some prerequisites from group theory.
\paragraph{Notations and conventions:} Throughout the present document, $\mathcal{M}$ denotes a time- and space-orientable spacetime with pseudo-Riemannian metric $g$ of signature $(-1, +1, +1, +1)$. $\eta$ will denote the Minkowski metric $\eta = \mathrm{diag}(-1, +1, +1, +1)$. We will denote the tangent bundle of $\mathscr{M}$ by $T\mathscr{M}$ and the tangent space at $p \in \mathscr{M}$ by $T_p\mathscr{M}$. Moreover, we will write $TU$ for the restriction of $T\mathscr{M}$ to an open set $U \subset \mathscr{M}$. The space of vector fields over an open set $U \subset \mathscr{M}$ will be written as $\mathfrak{X}(U)$ and $X \in \mathfrak{X}(\mathscr{M})$ denotes the global timelike vector field that describes the time orientation of $\mathscr{M}$. We assume in addition, that $X$ is normed, i.e. $g(X, X) = -1$, and denote the restriction of $X$ to $T_p\mathscr{M}$ by $X_p$. We denote the proper orthochronous Lorentz group by $\mathrm{SO}^+(1,3)$ and the group of all complex $2\times 2$ matrices with unit determinant by $\mathrm{SL}(2, \mathbb{C})$. Unit elements of matrix groups will be denoted by $\mathbf{1}$. Due to orientability of $\mathscr{M}$, the structure group of $T\mathscr{M}$ is reduced to $\mathrm{SO}^+(1,3)$, i.e. there exist bundle atlases for $T\mathscr{M}$ whose transition functions lie in $\mathrm{SO}^+(1,3)$. Let $\mathcal{C}$ be an open cover. We will then denote in the sequel by $\mathcal{A}= \{ (U, \psi)| U \in \mathcal{C} \}$ an atlas consisting out of local trivializations 
\begin{equation}
\psi: TU \rightarrow U \times \mathbb{R}^4
\end{equation}
that are induced by vielbein frames. I.e. for each $(U, \psi) \in \mathcal{A}$ there is an associated vielbein frame $(U, (E_\mu)_{\mu = 0, ..., 3})$ with $E_\mu \in \mathfrak{X}(U)$ satisfying
\begin{align}
g(E_\mu, E_\nu) &= \eta_{\mu \nu}\\
g(E_0, X) &= -1
\end{align}
such that for any $v = v^\mu E_\mu \in T_p\mathscr{M}$ with $p \in U$\begin{equation}
\psi(v^\mu E_\mu) = (p, (v^\mu) )
\end{equation}
holds. We will denote in the sequel vielbein frames just by $(U, E_\mu)$ or $(E_\mu)$. We will write the restriction of $g$ to $T_p\mathscr{M}$ as $g_p$. Let $p \in U$. The restriction of $(U, \psi)$ to $T_p\mathscr{M}$ with $p \in U$ will be written as
\begin{equation}
\psi_p: T_p\mathscr{M} \rightarrow \mathbb{R}^4, v^\mu E_\mu \mapsto (v^\mu),
\end{equation}
where $(E_\mu)$ is the vielbein associated to $(U, \psi)$. The restriction of $E_\mu$ to $p \in U$ will just be denoted by $E_\mu$. The euclidean norm on $\mathbb{R}^3$ will be denoted by $|\cdot|$. Moreover we define the $2$-sphere $S^2 \subset \mathbb{R}^3$ as
\begin{equation}
S^2 := \left\{ \hat{e} = (\hat{e}^1, \hat{e}^2, \hat{e}^3) \in \mathbb{R}^3 \middle| |\hat{e}| = 1\right\}
\end{equation}
and denote the Riemann sphere by $\mathbb{C}_\infty := \mathbb{C} \cup \{\infty\}$. Coordinate expressions in $\mathbb{R}^4$ will always be written as $(v^\mu)$ and vectors in $\mathbb{R}^3$ by $\vec{v}$ or $(v^i)$. Unit vectors in $\mathbb{R}^3$ will be denoted by $\hat{v}$. For a $4$-vector $(v^0, v^1, v^2, v^3)$ we define $\vec{v} := (v^1, v^2, v^3) = (v^i)$. All structures affiliated with the Riemann sphere will be introduced in the main text when they are needed and are additionally reviewed concisely in appendix \ref{LMCA1}. Finally, we define $\mathbb{R}^+ := (0, \infty)$ and denote the set of positive valued smooth functions on $\mathbb{C}_\infty$ by $C^\infty(\mathbb{C}_\infty, \mathbb{R}^+)$. Smooth functions associated with other domains are denoted analogously by $C^\infty(\cdot, \cdot)$. Products of smooth functions and numbers will be denoted by $\cdot$.

\section{Tangent light cones and their universal geometric structures}\label{TL}
In this section, we will introduce the basic geometric entity of our study, namely the bundle of future pointing light cones associated with a spacetime $\mathscr{M}$, and will analyse, which universal geometric structures are present on its fibers, as induced by the microscopic geometry of the generic spacetime $\mathscr{M}$. Therefore,  in section \ref{2.1} we will define the notion of tangent light cones as well as the light cone bundle. In addition, we will establish a bundle atlas for the light cone bundle, that is induced by the bundle atlas $\mathcal{A}$ of $T\mathscr{M}$ and is hence associated with vielbein frames. Moreover, we will sketch briefly, how transition functions look like for this bundle, and thereby, we will understand qualitatively how null vectors behave under Lorentz transformations. In section \ref{2.2} we will then investigate, which universal geometric structures on infinitesimal tangent light cones are induced by the microscopic geometry of $\mathscr{M}$, independently of the macroscopic behaviour of the metric $g$. 
\subsection{The light cone bundle}\label{2.1}
We define the pointed future\footnote{The analysis presented in this article goes completely analogous for past light cones and the associated past light cone bundle. Since $\mathscr{M}$ is time orientable, we focus hence without loss of generality solely on future light cones.} tangent light cone at $p \in \mathscr{M}$ as
\begin{equation}
L_p^+\mathscr{M} := \{ v \in T_p \mathscr{M} | g_p(v, v) = 0 \text{~and~} g(X_p, v) < 0 \}
\end{equation}
and by this, we define the future light cone bundle as the following sub-fiberbundle of $T\mathscr{M}$:
\begin{equation}
L^+\mathscr{M} := \bigsqcup_{p \in \mathscr{M}} L^+_p\mathscr{M} \subset T\mathscr{M}
\end{equation}
For any open set $U \subset \mathcal{M}$ we will denote the restriction of $L^+\mathscr{M}$ to $U$ by $L^+U$. We now introduce a class of suitable coordinate systems for $L^+\mathscr{M}$ which are induced by the bundle charts $(U, \psi) \in \mathcal{A}$ of $T\mathscr{M}$. We will then see especially, that the typical fiber of $L^+ \mathscr{M}$ is diffeomorphic to $\mathbb{C}_\infty \times \mathbb{R}^+$, i.e. for any $p \in \mathscr{M}$
\begin{equation}
L_p^+\mathscr{M} \cong \mathbb{C}_\infty \times \mathbb{R}^+
\end{equation}
holds. Here, $\mathbb{C}_\infty$ will be interpreted as a space of null directions and $\mathbb{R}^+$ as a "length" for null vectors. Therefore let $(U, \psi) \in \mathcal{A}$ be a bundle trivialization of $T\mathscr{M}$ and let $(E_\mu)$ be the associated vielbein. We then have for $v \in L_p^+\mathscr{M}$, that its coordinate representation $\psi_p(v) = (v^\mu)$ satisfies $v^0 = |\vec{v}|$, where we set $\vec{v} := (v^1, v^2, v^3)$. By this we can write $(v^\mu) = |\vec{v}| \cdot (1, \hat{v})$,
where we have defined $\hat{v}:= |\vec{v}|^{-1} \vec{v}$. Since $v^0 = |\vec{v}| > 0$ and $\hat{v} \in S^2$, we get therewith an identification
\begin{equation}
\tilde{\psi}_p^+: L^+_p \mathscr{M} \rightarrow  S^2 \times \mathbb{R}^+, v^\mu E_\mu \mapsto (\hat{v}, |\vec{v}|), \label{inm1}
\end{equation}
of the tangent lightcone at $p$ with $S^2 \times \mathbb{R}^+$. This gives then rise to a smooth bundle trivialization
\begin{equation}
\tilde{\psi}^+: L^+U \rightarrow U \times (S^2 \times \mathbb{R}^+), v^\mu E_\mu \in T_p\mathscr{M} \mapsto \left(p, (\hat{v}, |\vec{v}|)\right).
\end{equation}
We now utilize the stereographic projection defined as\footnote{We adopt the convention from \cite{Pen84}, which differs from the convention that is usually used in literature.}
\begin{equation}
\rho: S^2 \rightarrow \mathbb{C}_\infty, \hat{v} \mapsto \rho(\hat{v}) := \frac{\hat{v}^1 + i \hat{v}^2}{1 - \hat{v}^3}, 
\end{equation}
which constitutes a diffeomorphism. By this we get an identification
\begin{equation}
\psi^+_p: L_p^+\mathscr{M} \rightarrow \mathbb{C}_\infty \times \mathbb{R}^+,  v \mapsto \left(z^\psi_p(v), \lambda^\psi_p(v)\right) \label{rest}
\end{equation}
explicitely given by:
\begin{align}
z_p^\psi( v^\mu E_\mu) &:= \rho(\hat{v}) = \frac{v^1 + i v^2}{v^0 - v^3} \label{zd1} \\
\lambda_p^\psi(v^\mu E_\mu) &:= |\vec{v}| = v^0. \label{ld1}
\end{align}
This induces a smooth bundle trivialization
\begin{equation}
\psi^+: L^+U \rightarrow U \times (\mathbb{C}_\infty \times \mathbb{R}^+), v \in T_p \mathscr{M} \mapsto  \left(p, (z^\psi_p(v),  \lambda^\psi_p(v))  \right). \label{AM1}
\end{equation}
By this we have constructed a bundle atlas
\begin{equation}
\mathcal{B} := \{ (U, \psi^+) | (U, \psi) \in \mathcal{A} \}
\end{equation}
for the fiber bundle $L^+\mathscr{M}$, where for any $(U, \psi) \in \mathcal{A}$, the trivialization $\psi^+$ is given by the associated map \eqref{AM1}. For $p \in U$, the restriction of any trivialization $(U, \psi^+) \in \mathcal{B}$ to $L^+_p \mathscr{M}$ will be denoted by $\psi^+_p$ as given by \eqref{rest} and we will denote the set of all such restrictions by $\mathcal{B}_p$. Since Lorentz transformations preserve light cones, it is easy to show, that the transition functions of $\mathcal{B}$ are indeed smooth and preserve the fibers of $L^+\mathscr{M}$. Unfortunately, at the present stage we are not able to derive the precise form of the transition functions of $\mathcal{B}$ in a convenient representation. Nevertheless, we want to sketch already briefly the result, since it is illustrative for the understanding of the paper, although not absolutely necessary. The full discussion will then follow in sections \ref{EC} and \ref{GG} as well as in appendix \ref{LMCO3}. Hence, an impatient reader can also jump directly to the next subsection. Now let $p \in U$, set $v = v^\mu E_\mu \in L^+_p \mathscr{M}$ and $w = \Lambda^{\mu}_{~\nu} v^\nu  E_\mu$. Define in addition
\begin{align}
(z, \lambda) := \psi^+_p(v), \label{zl1} \\
(z', \lambda') := \psi^+_p(w). \label{zl2}
\end{align}
One can then show by utilization of the standard isomorphism between the Lorentz group $\mathrm{SO}^+(1,3)$ and the automorphism group of $\mathbb{C}_\infty$ (cf. appendix \ref{LMCO3} or \cite{Obl15, Pen84}), that there is a unique automorphism (i.e. a Möbius transformation, cf. appendix \ref{LMCA1}) $Z_\Lambda: \mathbb{C}_\infty \rightarrow \mathbb{C}_\infty$ of $\mathbb{C}_\infty$ associated to $\Lambda$ s.th.
\begin{equation}
z'= Z_\Lambda(z) \label{mt11}
\end{equation}
holds. By utilization of \eqref{ld1} we have naivly
\begin{equation}
\lambda'= \Lambda^{0}_{~\mu} v^\mu,
\end{equation}
which is not a very helpful representation, since it does not depend explicitely on $z$ and $\lambda$. But by some more advanced techniques (cf. appendix \ref{LMCO3}) one can indeed show, that there is for each Lorentz transformation $\Lambda \in \mathrm{SO}^+(1,3)$ an associated function $f_\Lambda \in C^\infty(\mathbb{C}_\infty, \mathbb{R}^+)$ such that
\begin{equation}
\lambda' = f_\Lambda(z) \lambda \label{resc11}
\end{equation}
holds. In this form, this is to the best of our knowledge an original result of this article. The concrete form of the function $f_\Lambda$ is not of importance now and will be understood in section \ref{EC}. At the present stage, just a qualitative understanding of equations \eqref{mt11} and \eqref{resc11} is sufficient: They state, that a Lorentz transformation acts on null vectors by a conformal transformation \eqref{mt11} on their space of directions $\mathbb{C}_\infty$, together with a non-trivial, direction dependent rescaling \eqref{resc11} of their "length". The latter can be understood by recalling, that
\begin{align}
\lambda = |\vec{v}| = v^0 \\
\lambda'= |\vec{w}| = w^0
\end{align}
hold  by (\ref{zl1} - \ref{zl2}) . We will understand later on, which non-trivial information is encoded in those rescalings.

\subsection{Universal structures on tangent light cones}\label{2.2}
We now want to analyze the geometric structures that are, independently of the macroscopic behaviour of the gravitational field, induced on tangent light cones $L_p^+\mathscr{M}$ by the geometry of their ambient tangent spaces $T_p\mathscr{M}$.  Those universal structures are linked to Einstein's equivalence principle, that is commonly stated as (cf. \cite{Car19}):
\begin{quote}
\textit{In small enough regions of spacetime, the laws of physics reduce to those of special relativity; it is impossible to detect the existence of a gravitational field by means of local experiments.}
\end{quote}
This statement should be understood as an \textit{asymptotic} statement: In the infinitesimal limit (and hence, strictly speaking, not locally\footnote{\label{fn1}Although "local" and "infinitesimal" are often used synonymous in the context of general relativity, there is a difference between those concepts: Infinitesimal objects are associated with the infinitesimal limit, i.e. with tangent spaces, while local objects are associated with (small) open sets and local coordinate systems. Those notions are hence not equivalent: For example, infinitesimally, one can always find vielbein frames in which the metric has Minkowski form, while it is not possible to find a local coordinate system with this property for a generic curved spacetime, cf. \cite{Poi04}. The latter can especially be formalized in terms of Riemann normal coordinates, cf. \cite{Poi04, Car19} and section \ref{6.1}.}), any spacetime behaves asymptotically like flat spacetime. Geometrically, this is formalized by the linear structure of any tangent space $T_p\mathscr{M}$ and by the property, that the metric $g$ reduces to an inner product $g_p$ on $T_p\mathscr{M}$ which can be brought to Minkowski form $\eta$ by choice of a vielbein frame. Hence the question of this section will be: What universal structures are induced on $L^+_p \mathscr{M}$ by the linear structure of $T_p\mathscr{M}$ and by the inner product $g_p$? And what are the coordinate expressions of those structures in the coordinate systems $\mathcal{B}_p$?

For the analysis of those questions, choose a $p \in \mathscr{M}$ and let $\psi^+_p \in \mathcal{B}_p$ be a coordinate system of the form \eqref{rest}. Observe then first, that $L_p^+\mathscr{M}$ inherits, as a subspace of the tangent space at $p$, a topology and a smooth structure, that can be equally characterized by the observation, that (\ref{rest}) is a diffeomorphism:
\begin{enumerate}[(S1)]
\item $L_p^+ \mathscr{M} \cong \mathbb{C}_\infty \times \mathbb{R}^+$ as a differentiable manifold.
\end{enumerate}
Moreover, although $L_p^+ \mathscr{M} \subset T_p \mathscr{M}$ is no linear subspace, it is a linear cone and as such it inherits a notion of multiplication with scalars from $T_p\mathscr{M}$:
\begin{enumerate}[(S2)]
\item $L^+_p \mathscr{M}$ is a linear cone, in the sense, that for all $\alpha > 0$  and all $v \in L^+_p \mathscr{M}$
\begin{equation}
\alpha \cdot v \in L^+_p \mathscr{M}
\end{equation}
holds. 
\end{enumerate}
Please notice at this point, that any coordinate system $\psi^+ \in \mathcal{B}$ respects this cone structure, since for all $\alpha > 0$ also
\begin{align}
z_p^\psi(\alpha \cdot v) &= z_p^\psi(v), \\
\lambda_p^\psi(\alpha \cdot v) &= \alpha\cdot \lambda_p^\psi (v)
\end{align}
hold. In addition, there is a kind of complex structure, which is induced by the complex structure on $\mathbb{C}_\infty$ under the identification of $L_p^+\mathscr{M}$ with $\mathbb{C}_\infty \times \mathbb{R}^+$:
\begin{itemize}[(S3)]
\item There is a set $\mathcal{Z}_p$ that includes exactly all surjective maps
\begin{equation}
z_p: L_p^+ \mathscr{M} \rightarrow \mathbb{C}_\infty 
\end{equation}
which satisfy the requirement, that
\begin{equation} 
 z_p^\psi \circ z_p^{-1}: \mathbb{C}_\infty \rightarrow \mathbb{C}_\infty
 \end{equation} 
is a biholomorphic automorphism of $\mathbb{C}_\infty$ and hence a Möbius transformation.
\end{itemize}
Please consult appendix \ref{LMCA1}, if you are not familiar with the notion of Möbius transformations. Finally we want to analyze, which structure on $L_p^+\mathscr{M}$ is induced by the metric $g$.  Prima facie, one could be tempted to think, that no interesting structure on $L_p^+\mathscr{M}$ is induced by the metric, since the latter is degenerate on thereon. But we will see, that this is not true. Therefore we will first introduce the inverse of the stereographic projection $\rho$, which will be denoted by $\hat{\epsilon}:= \rho^{-1}$, and is explicitely given by (cf. \cite {Pen84} or appendix \ref{LMCA1})
\begin{equation}
\hat{\epsilon}: \mathbb{C}_\infty \rightarrow S^2, z \mapsto \hat{\epsilon}(z) := (\hat{\epsilon}^1(z), \hat{\epsilon}^2(z), \hat{\epsilon}^3(z))
\end{equation}
with:
\begin{align}
\hat{\epsilon}^1(z) &= \frac{z + \bar{z}}{z \bar{z} + 1}, \\
\hat{\epsilon}^2(z) &= \frac{1}{i} \frac{z - \bar{z}}{z \bar{z} + 1}, \\
\hat{\epsilon}^3(z) &= \frac{z \bar{z} - 1}{z \bar{z} + 1}.
\end{align}
We will then write the inverse of $\psi^+_p$ for convenience as $\theta_p := (\psi^+_p)^{-1}$ and express it in terms of the inverse stereographic projection explicitely as
\begin{equation}
\theta_p: \mathbb{C}_\infty \times \mathbb{R}^+ \rightarrow L_p^+\mathscr{M}, (z, \lambda) \mapsto \lambda \cdot \hat{\epsilon}^\mu(z) E_\mu,
\end{equation}
where we set $\hat{\epsilon}^0 := 1$. By this we can then define a map
\begin{equation}
h_p: (\mathbb{C}_\infty \times \mathbb{R}^+) \times (\mathbb{C}_\infty \times \mathbb{R}^+) \rightarrow [0, \infty) \label{DF0}
\end{equation}
as the negative of the pullback of $g_p$ along $\theta_p$, i.e.:
\begin{equation}
h_p\left( (z_1, \lambda_1), (z_2, \lambda_2) \right) := - g_p(\theta_p(z_1, \lambda_1),\theta_p(z_2, \lambda_2) )
\end{equation}
As a somewhat surprising result we obtain then, that $h_p$ is explicitely given by
\begin{equation}
h_p\left( (z_1, \lambda_1), (z_2, \lambda_2) \right) = \frac{2 \lambda_1 \lambda_2 |z_1 - z_2|^2}{(|z_1|^2 + 1) (|z_2|^2 +1)},
\end{equation}
which is just
\begin{equation}
h_p\left( (z_1, \lambda_1), (z_2, \lambda_2) \right)  = \frac{\lambda_1 \lambda_2}{2} d^2(z_1, z_2) \label{DF1}
\end{equation}
with $d$ being the chordal distance (cf. \eqref{CDApp} and \eqref{CDApp2}) on $\mathbb{C}_\infty$ given by:
\begin{equation}
d: \mathbb{C}_\infty \times \mathbb{C}_\infty \rightarrow [0 ,\infty) , (z_1, z_2) \mapsto \frac{2 |z_1 - z_2|}{\sqrt{|z_1|^2 + 1} \sqrt{|z_2|^2 + 1}}. \label{CD1}
\end{equation} 
Hence, although the metric is degenerate on $L_p^+\mathscr{M}$, it carries non-trivial information, since it describes a kind of distance function\footnote{Please note, that neither $h_p$ nor $\sqrt{h_p}$ constitute (pseudo-)metrics, since they do not obey the triangle equality. Nevertheless, they can be interpreted as distance functions in the present situation, since they are induced by a degenerate Riemannian metric on $L_p^+\mathscr{M}$ and are related to the chordal distance on $\mathbb{C}_\infty$.  Therefore we call them distance functions in the sequel.} (\ref{DF1}) thereon, which is related to the chordal distance on $\mathbb{C}_\infty$. We now can ask, if the distance function (\ref{DF1}) is related to some kind of Riemannian metric on the manifold $\mathbb{C}_\infty \times \mathbb{R}^+$. And indeed, one can show easily, that $\sqrt{h_p}$ is the distance function induced by a degenerate Riemannian metric $q_p$ on $L_p^+\mathscr{M}$, which is given in any coordinate system $\psi_p^+ \in \mathcal{B}_p$ by:
\begin{equation}
ds^2  =2 \lambda^2 \frac{dz d\bar{z}}{(1 + z \bar{z})^2} \label{mc1}
\end{equation}
I.e. we have:
\begin{enumerate}[(S4)]
\item There exists a degenerate metric $q_p$ on $L_p^+\mathscr{M}$ whose coordinate expression in any coordinate system $\psi^+_p \in \mathcal{B}_p$ is given by
\begin{equation}
ds^2  = 2 \lambda^2 \frac{dz d\bar{z}}{(1 + z \bar{z})^2}.
\end{equation}
\end{enumerate}
Please note, that $q_p$ is really a degenerate Riemannian metric on the manifold $L_p^+\mathscr{M}$, as (\ref{mc1}) is a degenerate Riemannian metric on the manifold $\mathbb{C}_\infty \times \mathbb{R}^+$. This is in contrast to the metric $g$, which is a pseudo-Riemannian metric on $\mathscr{M}$ and gives as such rise to an inner product $g_p$ on $T_p^+\mathscr{M}$. As such, $g_p$ induces then the distance function (\ref{DF1}) on $L_p^+\mathscr{M} \subset T_p\mathscr{M}$, whose infinitesimalization is then given by (\ref{mc1}), which describes $q_p$ in the coordinates $\psi^+_p$. 

Hence we have all together the following universal structures on $L_p^+\mathscr{M}$:
\begin{enumerate}[(S1)]
\item $L_p^+ \mathscr{M} \cong \mathbb{C}_\infty \times \mathbb{R}^+$ as a differentiable manifold. \label{S1}
\item $L^+_p \mathscr{M}$ is a linear cone, in the sense that for all $\alpha > 0$ and all $v \in L_p^+\mathscr{M}$ 
\begin{equation}
\alpha \cdot v \in L^+_p \mathscr{M}
\end{equation}
holds. \label{S2}
\item There is a set $\mathcal{Z}_p$ which includes exactly all surjective maps
\begin{equation}
z_p: L_p^+ \mathscr{M} \rightarrow \mathbb{C}_\infty 
\end{equation}
that satisfy the requirement, that
\begin{equation} 
 z_p^\psi \circ z_p^{-1}: \mathbb{C}_\infty \rightarrow \mathbb{C}_\infty
 \end{equation} 
is biholomorphic for any coordinate system $\psi^+_p = (z_p^\psi, \lambda_p^\psi) \in \mathcal{B}_p$ and hence a Möbius transformation. \label{S3}
\item There exists a degenerate metric $q_p$ on $L_p^+\mathscr{M}$, whose expression in any local coordinate system $\psi_p^\psi \in \mathcal{B}_p$ is given by the metric $\tilde{q}_p$ on $\mathbb{C}_\infty \times \mathbb{R}^+$ explicitely given by
\begin{equation}
ds^2  = 2 \lambda^2 \frac{dz d\bar{z}}{(1 + z \bar{z})^2}. \label{mets}
\end{equation} \label{S4}
\end{enumerate}
In the next section we want to analyze the set of automorphisms of $L_p^+\mathscr{M}$, which preserve those universal geometric structures.  

\section{Automorphism groups of tangent light cones}\label{4}
In this section we will determine the automorphisms of $L_p^+\mathscr{M}$ which preserve the universal structures \ref{S1} - \ref{S4} in a certain sense. As explained in the introduction, the crucial question is how to interpret the structure \ref{S4}: Should it be understood as a single degenerate metric or as a representative of a conformal equivalence class on $L_p^+\mathscr{M} \cong \mathbb{C}_\infty \times \mathbb{R}^+$? The former will lead to a group of isometries, while the latter will give a conformal automorphism group. There are arguments for both positions and we will comment on this more extensively in sections \ref{OLG} and \ref{Conc}. But in this section, we won't bother with this question and just determine the respective automorphism groups for both interpretations. Therefore we will explain some generalities regarding automorphism groups of $L_p^+\mathscr{M}$ in section \ref{4.1} and will especially analyse there, how the structures \ref{S1} - \ref{S3} already fix the form of suitable automorphisms to a great extent, independent of the interpretation of \ref{S4}. Moreover, we will review in this subsection some basics regarding Möbius transformations and metrics on the Riemann sphere, which are needed in the sequel. In section \ref{4.2} we will then determine the group of isometries, while in section \ref{4.3} we will determine the conformal automorphism group. 
\subsection{Generalities}\label{4.1}
An automorphism group of $L_p^+\mathscr{M}$ is a set of all maps
\begin{equation}
\Pi_p: L_p^+\mathscr{M} \rightarrow L_p^+\mathscr{M}
\end{equation}
which preserve the universal structures \ref{S1} - \ref{S4} in a specific sense, together with the composition $\circ$ as a group operation. By choosing an arbitrary but fixed coordinate system $\psi_p^+ \in \mathcal{B}_p$, one notices, that a map $\Pi_p$ preserves the structures \ref{S1} - \ref{S4} in a specified sense if and only if the associated coordinate representation of $\Pi_p$ given by
\begin{equation}
 \psi^+_p \circ \Pi_p \circ \theta_p: \mathbb{C}_\infty \times \mathbb{R}^+ \rightarrow \mathbb{C}_\infty \times \mathbb{R}^+
\end{equation}
preserves the induced universal structures on $\mathbb{C}_\infty \times \mathbb{R}^+$ in the analogous specified sense. Here we set again for convenience $\theta_p := (\psi_p^+)^{-1}$. Additionally, the group multiplication law $\circ$ is preserved under conjugation with $\psi^+_p$  and hence any automorphism group of $L_p^+\mathscr{M}$ is naturally isomorphic to the corresponding automorphism group of the induced structures on $\mathbb{C}_\infty \times \mathbb{R}^+$. Hence, we can define the automorphism group $L_p^+\mathscr{M}$ equally as a group of automorphisms of $\mathbb{C}_\infty \times \mathbb{R}^+$ together with $\circ$ as a group operation and this is what we will do in the sequel, since it is more convenient in the present situation.

As said before, there are two different ways, how one could interpret the structure \ref{S4}: Either as a degenerate metric on $L_p^+\mathscr{M}$ or as a representative of a conformal equivalence class  thereon. Each of those two interpretations yields then its respective automorphism group. In the former case, the automorphisms constitute isometries of $L_p^+\mathscr{M}$, while in the latter case they correspond to conformal automorphisms of $L_p^+\mathscr{M}$. Hence, we will obtain two distinct automorphism groups: The group of isometries $\mathrm{Iso}^+_p$ of $L_p^+\mathscr{M}$ and the group of conformal automorphisms $\mathrm{Con}^+_p$ of $L_p^+\mathscr{M}$.
Nevertheless, both automorphism groups should preserve the universal structures \ref{S1} - \ref{S3} in the same sense, since those structures are independent of the interpretation of \ref{S4}. Therefore, we will discuss now first the requirements on automorphisms as induced by \ref{S1}- \ref{S3}. For this, we write first a generic map 
\begin{align}
\Phi_p:~ &\mathbb{C}_\infty \times \mathbb{R}^+ \rightarrow \mathbb{C}_\infty \times \mathbb{R}^+ \label{Au2}
\end{align}
in full generality as
\begin{align}
\Phi_p: (z, \lambda) \mapsto (Z(z, \lambda), Y(z, \lambda)).
\end{align}
We then have first and foremost the following two requirements for $\Phi_p$, which are induced by \ref{S1} and \ref{S2} respectively:
\begin{enumerate}[(R1)]
\item $\Phi_p$ is a diffeomorphism. \label{R1}
\item $\Phi_p$ is homogeneous in the sense, that for all $\alpha > 0$ and all $(z, \lambda) \in \mathbb{C}_\infty \times \mathbb{R}^+$
\begin{align}
Z(z, \alpha \cdot \lambda) &= Z(z, \lambda)  \label{impl1}\\
Y(z, \alpha \cdot \lambda) &= \alpha \cdot Y(z, \lambda) \label{impl2}
\end{align}
should hold. \label{R2}
\end{enumerate}
Those both requirements fix the form of $\Phi_p$ already to a great extent: Equation (\ref{impl1}) and requirement \ref{R1} are together equivalent to the statement, that $Z$ is a smooth automorphism of $\mathbb{C}_\infty$ which does not depend on the $\lambda$-coordinate. Moreover, equation (\ref{impl2}) and \ref{R2} hold together if and only if there exists a smooth function $Y \in C^\infty(\mathbb{C}_\infty, \mathbb{R}^+)$ such that
\begin{equation}
Y(z, \lambda) = Y(z) \cdot \lambda
\end{equation}
holds for all $(z, \lambda) \in \mathbb{C}_\infty \times \mathbb{R}^+$. Consequently, we can write any map (\ref{Au2}) satisfying the requirements \ref{R1} - \ref{R2} as
\begin{align}
\Phi_p:~ &\mathbb{C}_\infty \times \mathbb{R}^+ \rightarrow \mathbb{C}_\infty \times \mathbb{R}^+, (z, \lambda) \mapsto (Z(z), Y(z) \cdot \lambda)
\end{align}
with $Y \in C^\infty(\mathbb{C}_\infty, \mathbb{R}^+)$ and $Z$ being a diffeomorphism $Z: \mathbb{C}_\infty \rightarrow \mathbb{C}_\infty$.
Now, the complex structure \ref{S3} is preserved if and only if $Z$ is a biholomorphic conformal automorphism of $\mathbb{C}_\infty$, i.e. a Möbius transformation. This gives the following requirement:
\begin{enumerate}[(R3)]
\item $Z$ is a biholomorphic automorphism of the Riemann sphere $\mathbb{C}_\infty$, i.e. a Möbius transformation. \label{R3}
\end{enumerate}
All together, any automorphism of $\mathbb{C}^\infty \times \mathbb{R}^+$ which preserves the structures \ref{S1} - \ref{S3} and satisfies consequently the requirements \ref{R1} - \ref{R3} is given by a map
\begin{align}
\Phi_p:~ &\mathbb{C}_\infty \times \mathbb{R}^+ \rightarrow \mathbb{C}_\infty \times \mathbb{R}^+, (z, \lambda) \mapsto (Z(z), Y(z) \cdot \lambda) \label{AuFor}
\end{align}
with $Z$ being a Möbius transformation and $Y \in C^\infty(\mathbb{C}_\infty, \mathbb{R}^+)$. Moreover we see, that two automorphisms are equal if and only if their associated Möbius transformations and smooth functions $Y$ are equal, respectively.

In the sequel we will need some basic facts regarding Möbius transformations. Hence, we will recapitulate those facts concisely now, while a more complete account of the corresponding theory is presented in appendix \ref{LMCA1}. Therefore recall first, that any Möbius transformation $Z$ can be written as
\begin{equation}
Z: \mathbb{C}_\infty, \rightarrow \mathbb{C}_\infty, z \mapsto  \frac{a z + b}{cz + d}\label{MT1}
\end{equation}
for complex numbers $a,b,c,d \in \mathbb{C}$ satisfying $ad -bc = 1$. By this, any matrix $A \in \mathrm{SL}(2, \mathbb{C})$ given by
\begin{equation}
A = \begin{pmatrix} a & b \\ c & d \end{pmatrix} \label{MZ1}
\end{equation}
defines an associated Möbius transformation $Z^A$ specified by
\begin{equation}
Z^A(z) := \frac{az + b}{cz +d}. \label{MZ2}
\end{equation}
Moreover, two matrices $A, B \in \mathrm{SL}(2, \mathbb{C})$ define the same Möbius transformation if and only if $A = -B$. This defines then an isomorphism between $\mathrm{PSL}(2, \mathbb{C}) := \mathrm{SL}(2, \mathbb{C}) / \{ \pm \mathbf{1} \}$  and the Möbius group, that is explicitely given by
\begin{equation}
[A] \in \mathrm{PSL}(2, \mathbb{C}) \mapsto Z^A
\end{equation}
with $Z^A$ as defined in \eqref{MZ2}. We will use in the sequel the notions Möbius group and $\mathrm{PSL}(2, \mathbb{C})$ interchangeably. In addition, we will denote equivalence classes $[A] \in \mathrm{PSL}(2, \mathbb{C})$ just in terms of one of their representatives, i.e. by slight abuse of notation $[A] = A$. Now recall further, that $\mathbb{C}_\infty$ is a Riemann surface and is hence especially endowed with a natural conformal structure (cf. \cite{Jos06}), that can be described in terms of a conformal equivalence class of Riemannian metrics. One representative of this conformal structure is given by the Riemannian metric
\begin{equation}
ds^2 = \frac{4}{(1 + z \bar{z})^2} dz d\bar{z}
\end{equation}
on $\mathbb{C}_\infty$, which is also the Riemannian metric associated with the chordal distance \eqref{CD1}. The pullback of this metric along a Möbius transformation $Z^A$ as specified by \eqref{MZ2} is then explicitely given by
\begin{equation}
ds^2 = K^A(z)^2 \frac{4}{(1 + z \bar{z})^2} dz d\bar{z}, \label{PMB1}
\end{equation}
where the conformal factor $K^A(z)$ associated with a matrix $A \in \mathrm{PSL}(2, \mathbb{C})$ as specified by \eqref{MZ1} is explicitely given by
\begin{equation}
K^A(z) = \frac{(1 + z \bar{z})}{(az + b) (\bar{a} \bar{z} + \bar{b})+ (c z + d) (\bar{c} \bar{z} + \bar{d})}, \label{KF1}
\end{equation}
as one can easily calculate. Finally recall, that the Möbius group and the proper orthochronous Lorentz group are isomorphic, i.e. $\mathrm{PSL}(2, \mathbb{C}) \cong \mathrm{SO}^+(1,3)$, as said before and as described in appendix \ref{LMCO3} (cf. also \cite{Pen84, Obl15}). The corresponding isomorphism will be written as
\begin{equation}
\Lambda \in \mathrm{SO}^+(1,3) \mapsto A_\Lambda \in \mathrm{PSL}(2, \mathbb{C}).
\end{equation}
\subsection{The isometry group}\label{4.2}
We now want to find all maps
\begin{align}
\Phi_p:~ &\mathbb{C}_\infty \times \mathbb{R}^+ \rightarrow \mathbb{C}_\infty \times \mathbb{R}^+, (z, \lambda) \mapsto (Z(z), Y(z) \cdot \lambda)
\end{align}
of the form (\ref{AuFor}), such that the following requirement holds:
\begin{enumerate}[(R4a)]
\item $\Phi_p$ is an isometry, i.e.
\begin{equation}
\Phi_p^* 	\tilde{q} _p = \tilde{q}_p
\end{equation}
is satisfied, where $\tilde{q}_p$ is the metric (\ref{mets}) on $\mathbb{C}_\infty \times \mathbb{R}^+$ that is induced by the metric $q_p$ on $L^+_p \mathscr{M}$, cf. \ref{S4}. \label{R4a}
\end{enumerate} 
Recall from \ref{S4}, that $\tilde{q}_p$ is given by:
\begin{equation}
ds^2 = 2 \lambda^2 \frac{dz d\bar{z}}{(1 + z \bar{z})^2}
\end{equation}
Then its pullback $\Phi_p^* \tilde{q} _p$ can be easily calculated by the utilization of \eqref{PMB1} and is explicitely given by
\begin{equation}
ds^2 = 2 Y(z)^2 \lambda^2 K^A(z)^2 \frac{dz d\bar{z}}{(1 + z \bar{z})^2}, \label{PBM2}
\end{equation}
where $K^A$ is the conformal factor (\ref{CF}). Hence we obtain directly, that $\Phi_p$ is an isometry if and only if
\begin{align}
Y(z) \overset{!}{=} K^A(z)^{-1}
\end{align}
holds. We  now define 
\begin{equation}
f^A := (K^A)^{-1} \label{fadef}
\end{equation}
and by this
\begin{equation}
Y(z) \overset{!}{=} f^A(z)
\end{equation}
should hold for $\Phi_p$ being an isometry. By this we have, that the isometry group of the infinitesimal light cone consists out of all maps $\Phi_p^A$ of the form
\begin{align}
\Phi_p^A : \mathbb{C}_\infty \times \mathbb{R}^+ \rightarrow \mathbb{C}_\infty \times \mathbb{R}^+, (z, \lambda) \mapsto (Z_A(z), f^A(z) \lambda),
\end{align}
where $A$ is any matrix $A \in \mathrm{PSL}(2, \mathbb{C})$, together with the composition as the group multiplication law. This group will be denoted in the sequel as:
\begin{equation}
\mathrm{Iso}^+_p = \left( \{ \Phi_p^A | A \in \mathrm{PSL}(2, \mathbb{C})\}, \circ \right).
\end{equation}
We see already, that it is isomorphic to $\mathrm{PSL}(2, \mathbb{C})$ as a set, and hence also isomorphic to the Lorentz group $\mathrm{SO}^+(1,3)$ as a set. We will understand in section \ref{EC}, that they are also isomorphic as a group, and that $\mathrm{Iso}^+_p$ is induced by local Lorentz transformations acting on $T_p\mathscr{M}$. By this, it is also not a coincidence, that the function \eqref{fadef} and the function $f_\Lambda$ which determines the rescalings \eqref{resc11} of null vectors under Lorentz transformations are both denoted by the same letter. Indeed, they are the same mathematical object, as we will understand then, too.
\subsection{The conformal automorphism group}\label{4.3}
We now want to find all maps
\begin{align}
\Psi_p:~ &\mathbb{C}_\infty \times \mathbb{R}^+ \rightarrow \mathbb{C}_\infty \times \mathbb{R}^+, (z, \lambda) \mapsto (Z(z), Y(z) \cdot \lambda)
\end{align}
of the form (\ref{AuFor}), such that the following requirement holds:
\begin{enumerate}[(R4b)]
\item $\Psi_p$ is a conformal automorphism, i.e. there exists a function $\Omega_p \in C^\infty(\mathbb{C}_\infty \times \mathbb{R}^+, \mathbb{R}^+)$ s.th.
\begin{equation}
\Psi_p^* \tilde{q}_{p, (z, \lambda)} = \Omega_p(z, \lambda)^2 \tilde{q}_{p,(z,\lambda)} \label{CFEe}
\end{equation}
holds for all $(z, \lambda) \in \mathbb{C}_\infty \times \mathbb{R}^+$. Here $\tilde{q}_p$ denotes,  as before, the metric (\ref{mets}) on $\mathbb{C}_\infty \times \mathbb{R}^+$ and $\tilde{q}_{p, (z, \lambda)}$ denotes its pointwise evaluation on the tangent space $T_{(z, \lambda)} (\mathbb{C}_\infty \times \mathbb{R}^+)$ of $\mathbb{C}_\infty \times \mathbb{R}^+$ at $(z, \lambda)$. \label{R4b}
\end{enumerate}
Observe, that, due to the form (\ref{AuFor}) of $\Psi_p$, it follows directly, that the conformal factor must be of the form $\Omega_p(z)$ with $\Omega_p \in C^\infty(\mathbb{C}_\infty, \mathbb{R}^+)$. We have namely as in (\ref{PBM2})
\begin{equation}
\Psi_p^* \tilde{q}_{p, (z, \lambda)} = Y(z)^2 K^A(z)^2 \tilde{q}_{p, (z, \lambda)},
\end{equation}
which says, that the conformal factor $\Omega_p$ as defined in \eqref{CFEe} is explicitely given by:
\begin{equation}
\Omega_p(z) = Y(z) K^A(z)
\end{equation}
By this, the group of conformal automorphisms of the infinitesimal light cone consists out of all maps
\begin{align}
\Psi_p^{(A, Y)}: &\mathbb{C}_\infty \times \mathbb{R}^+ \rightarrow \mathbb{C}_\infty \times \mathbb{R}^+, (z, \lambda) \mapsto (Z_A(z), Y(z) \cdot \lambda),
\end{align}
with $A \in \mathrm{PSL}(2, \mathbb{C})$ and $Y \in C^\infty(\mathbb{C}_\infty, \mathbb{R}^+)$, together with the composition as the group multiplication law. This group will be denoted in the sequel as:
\begin{equation}
 \mathrm{Con}_p^+ := \left( \{ \Psi_p^{(A, Y)} | A \in \mathrm{PSL}(2, \mathbb{C}) \text{~and~} Y \in C^\infty(\mathbb{C}_\infty, \mathbb{R}^+)\}, \circ \right).
\end{equation}
We see already, that it is isomorphic to $\mathrm{PSL}(2, \mathbb{C}) \times C^\infty(\mathbb{C}_\infty, \mathbb{R}^+)$ as a set. Its non-trivial group structure will be understood in section \ref{5}.

We want to remark finally, that the condition \ref{R4b} can be rephrased in a different but equivalent manner. Therefore observe first, that the metric $\tilde{q}_p$ from (\ref{mets}) is compatible with the linear cone structure \ref{S2} of $L_p^+\mathscr{M}$ in the sense, that
\begin{equation}
\tilde{q}_{p, (z, \alpha \lambda)} = \alpha^2 \tilde{q}_{q, (z, \lambda)}
\end{equation}
holds for all $\alpha > 0$. This behaviour is induced by the compatibility of the distance function $h_p$ (\ref{DF0}) with the cone structure. If one would like to define a conformal equivalence class of metrics on $L_p^+\mathscr{M}$, all metrics in this class should have this property, because otherwise they would not be compatible with \ref{S2}. Hence, two metrics $q^{(1)}_p$ and $q^{(2)}_p$ on $L_p^+\mathscr{M}$ should be called conformally equivalent, if there exists a positive valued, smooth function $\Omega_p \in C^\infty(\mathbb{C}_\infty, \mathbb{R}^+)$ independent of $\lambda$, such that their coordinate expressions $\tilde{q}^{(1)}_p, \tilde{q}^{(2)}_p$ satisfy:
\begin{equation}
\tilde{q}^{(1)}_{p, (z, \lambda)} = \Omega_p(z)^2 \tilde{q}^{(2)}_{p, (z, \lambda)} 
\end{equation}
The conformal structure on $L_p^+\mathscr{M}$ is then the set of all metrics that are in this sense conformally equivalent to the metric (\ref{mets}). One can then see easily, that the requirement \ref{R4b} is, due to the requirement \ref{R2}, equivalent to the statement, that any conformal automorphism of $L_p^+\mathscr{M}$ should preserve this conformal structure under pullback.
\subsection{Summary}\label{4.4}
In this section we have found two natural automorphism groups for $\mathrm{L}^+_p\mathscr{M}$. The first is the group of isometries $\mathrm{Iso}^+_p$, whose underlying set is given by
\begin{equation}
\mathrm{Iso}^+_p = \{ \Phi_p^{A} | A \in \mathrm{PSL}(2, \mathbb{C})\}
\end{equation}
with
\begin{align}
\Phi_p^A : \mathbb{C}_\infty \times \mathbb{R}^+ \rightarrow \mathbb{C}_\infty \times \mathbb{R}^+, (z, \lambda) \mapsto (Z_A(z), f^A(z) \lambda),
\end{align}
together with the composition $\circ$ as the group operation. Here $f^A(z)$ is defined by \eqref{fadef}. The other is the group of conformal automorphisms $\mathrm{Con}_p^+$, whose underlying set is given by
\begin{equation}
\mathrm{Con}_p^+ = \{ \Psi_p^{(A, Y)} | A \in \mathrm{PSL}(2, \mathbb{C}), Y \in C^\infty(\mathbb{C}_\infty, \mathbb{R}^+)\}
\end{equation}
with
\begin{align}
\Psi_p^{(A, Y)}:\mathbb{C}_\infty \times \mathbb{R}^+ \rightarrow \mathbb{C}_\infty \times \mathbb{R}^+, (z, \lambda) \mapsto (Z_A(z), Y(z) \cdot \lambda), \label{PDF}
\end{align}
together with the composition $\circ$ as the group operation. In the next section we will understand their mathematical structure and their interrelation.
\section{Structure of the automorphism groups}\label{5}
In the last section we have determined two different automorphism groups of $L_p^+\mathscr{M}$: A isometry group $\mathrm{Iso}^+_p$ and a conformal automorphism group $\mathrm{Con}^+_p$. In this section, we want to understand their structure, interrelation and interpretation better. Therefore we will analyze the mathematical structure of the conformal automorphism group $\mathrm{Con}^+_p$ in section \ref{5.1} and will show, that it constitutes a semidirect product of the Möbius group with the group of positive valued smooth functions on the Riemann sphere. In section \ref{5.2} we will then show, that the conformal automorphism group contains infinitely many Lorentz subgroups, and explain, how those subgroups can be parametrized in terms of so called crossed homomorphisms. In section \ref{5.3} we will explain, how the isometry group $\mathrm{Iso}^+_p$ is related to the conformal automorphism group $\mathrm{Con}^+_p$. Especially we will show there, that $\mathrm{Iso}^+_p$ constitutes a specific, non-canonical Lorentz subgroup of $\mathrm{Con}^+_p$ which is induced by the representation of the Lorentz group on $T_p \mathscr{M}$. In section \ref{OLG} we will then try to make the structure and interpretation of $\mathrm{Con}^+_p$ more lucid by explaining, how any Lorentz subgroup of $\mathrm{Con}^+_p$ seems to set a notion of scale for null vectors. By this we will argue, that $\mathrm{Con}^+_p$ could be a more natural automorphism group for $L_p^+\mathscr{M}$ than any of its Lorentz subgroups. In the whole section we will need some prerequisites from group theory as adapted to our situation. Especially the notions of right semidirect products and crossed homomorphisms are needed. To increase readability, we have outsourced their discussion to appendix \ref{BGT}. Please note, that we will denote throughout this section the element in $\mathrm{PSL}(2, \mathbb{C})$ associated with a Lorentz transformation $\Lambda \in \mathrm{SO}^+(1,3)$ by $A_\Lambda \in \mathrm{PSL}(2, \mathbb{C})$ and the corresponding Möbius transformation by $Z_\Lambda := Z^{A_\Lambda}$.
\subsection{Structure of the conformal automorphism group}\label{5.1}
Let $G$ be the (right) semidirect product group (for a definition of this notion see appendix \ref{BGT1})
\begin{equation}
 G := \mathrm{PSL}(2, \mathbb{C}) \ltimes_\kappa C^\infty(\mathbb{C}_\infty, \mathbb{R}^+),
 \end{equation}
 where the group antihomomorphism
 \begin{equation}
 \kappa: \mathrm{PSL}(2, \mathbb{C}) \rightarrow \mathrm{Aut}(C^\infty(\mathbb{C}_\infty, \mathbb{R}^+)), A \mapsto \kappa_A \label{AH1}
 \end{equation}
is defined as:
 \begin{equation}
 \kappa_A: C^\infty(\mathbb{C}_\infty, \mathbb{R}^+) \rightarrow C^\infty(\mathbb{C}_\infty, \mathbb{R}^+), Y \mapsto Y \circ Z^A \label{AH2}
\end{equation}
This means in particular, that $G \cong \mathrm{PSL}(2, \mathbb{C}) \times C^\infty(\mathbb{C}_\infty, \mathbb{R}^+)$ as a set and that the product of $(A_1, Y_1), (A_2, Y_2) \in G$ is defined as:
\begin{equation}
(A_1, Y_1) (A_2, Y_2) = (A_1 A_2, Y_1 \circ Z^{A_2} \cdot Y_2). \label{PG1s}
\end{equation}
We will show in this section, that the group $\mathrm{Con}^+_p$ of conformal automorphisms is isomorphic to $G$ and moreover, that the action of $\mathrm{Con}^+_p$ on $\mathbb{C}_\infty \times \mathbb{R}^+$ can be described in terms of a faithful left group action $\star: G \curvearrowright \mathbb{C}_\infty \times \mathbb{R}^+$. The latter means, that $G$ is indeed a transformation group (cf. \cite{Ham17}) acting on $\mathbb{R}^+ \times \mathbb{C}_\infty$. 

Therefore observe first, that obviously $\mathrm{Con}^+_p$ and $G$ are isomorphic as sets in terms of the bijection
\begin{equation}
(A, Y) \in G \mapsto \Psi^{(A,Y)},
\end{equation}
with $\Psi^{(A,Y)} \in \mathrm{Con}^+_p$ being defined as in (\ref{PDF}). Let now $(A_1, Y_1), (A_2, Y_2) \in G$. Then their product in $G$ is given by \eqref{PG1s}, while we have on the other hand:
\begin{align}
\Psi^{(A_1, Y_1) } \circ \Psi^{(A_2, Y_2)} &= \Psi_p^{(A_1A_2, Y_1 \circ Z^{A_2}\cdot Y_2)}.
\end{align}
Hence the map
\begin{equation}
G \rightarrow \mathrm{Con}^+_p, g \mapsto \Psi_p^g
\end{equation}
is indeed a group isomorphism. Define now a faithful left group action of $G$ on $\mathbb{C}_\infty \times \mathbb{R}^+$
\begin{equation}
\star: G \curvearrowright \mathbb{C}_\infty \times \mathbb{R}^+, (g, (z, \lambda)) \mapsto g \star (z, \lambda) \label{star}
\end{equation} 
as:
\begin{equation}
(A, Y) \star (z, \lambda) = (Z^A(z), Y(z) \lambda).
\end{equation}
Let now $(z, \lambda) \in \mathbb{C}_\infty \times \mathbb{R}^+$. Then one can show easily:
\begin{equation}
\forall g \in G: \Psi_p^g(z, \lambda) = g \star (z, \lambda).
\end{equation}
Hence we have shown, that $\mathrm{Con}^+_p$ is isomorphic to $G$ and acts on $\mathbb{C}_\infty \times \mathbb{R}^+$ in terms of the faithful group action $\star$. Therefore we will use the notions $G$ and $\mathrm{Con}^+_p$ synonymously, while we prefer the former in situations, where we focus on the mathematical structure of the conformal automorphism group, and use the latter, when we refer to its interpretation as a group of automorphisms of $L_p^+\mathscr{M}$.
\subsection{Lorentz subgroups in terms of crossed homomorphisms}\label{5.2}
Let now
\begin{equation}
c: \mathrm{SO}^+(1,3) \rightarrow C^\infty(\mathbb{C}_\infty, \mathbb{R}^+), \Lambda \mapsto c_\Lambda
\end{equation}
be a crossed homomorphism (see appendix \ref{BGT2}), i.e. $c$ satisfies
\begin{equation}
c_{\Lambda_1 \Lambda_2} = c_{\Lambda_2} \cdot c_{\Lambda_1}\circ Z_{\Lambda_2}.
\end{equation}
By isomorphy of the Lorentz group $\mathrm{SO}^+(1,3)$ and the Möbius group $\mathrm{PSL}(2, \mathbb{C})$, this induces a Lorentz subgroup $i_c(\mathrm{SO}^+(1,3)) \subset G$ as specified by the associated embedding
\begin{equation}
i_c: \mathrm{SO}^+(1,3) \hookrightarrow G, \Lambda \mapsto (A_\Lambda, c_\Lambda ), \label{emb11}
\end{equation}
and moreover, any Lorentz subgroup of $G$ is of this form. This is a general property of semidirect product groups, as explained in appendix \ref{BGT2}. We now want to give two examples for classes of crossed homomorphisms in the present situation. For both classes, there exist elementary examples, that we have encountered already without noticing it or that occur later on. A complete classification of crossed homomorphisms is not important for the present discussion and will be a question of further research, cf. section \ref{OOQ}. For the first class, let $L \in C^\infty(\mathbb{C}_\infty, \mathbb{R}^+)$ be a smooth function. It is then easy to see, that the associated map
\begin{equation}
c^{(L)}: \mathrm{SO}^+(1,3) \rightarrow C^\infty(\mathbb{C}_\infty, \mathbb{R}^+), \Lambda \mapsto \frac{L \circ Z_\Lambda}{L}
\end{equation}
constitutes a crossed homomorphism and hence, the associated embedding
\begin{equation}
\mathrm{SO}^+(1,3) \hookrightarrow G, \Lambda \mapsto \left(A_\Lambda,  \frac{L \circ Z_\Lambda}{L}\right)
\end{equation}
defines a Lorentz subgroup of $G$. For the second class of crossed homomorphisms, we have to introduce so called projective coordinates for $\mathbb{C}_\infty$ (cf. appendix \ref{LMCA1} or \cite{Pen84}). Those are obtained by labeling points $z \in \mathbb{C}_\infty$ not by a single complex number,  but by a pair of complex numbers $(\xi, \eta) \in \mathbb{C}^2$ that are allowed to take any value other than $(0,0)$ and specify a point $z \in \mathbb{C}_\infty$ in terms of the quotient
\begin{equation}
z = \xi / \eta.
\end{equation}
They are often more convenient, since any point on $\mathbb{C}_\infty$ can be labeled in terms of projective coordinates by two finite complex numbers, e.g. $\infty = \xi/0$. Also, the action of Möbius transformations on projective coordinates can be conveniently expressed (by slight abuse of notation) as:
\begin{equation}
Z^A(\xi/\eta) = A \begin{pmatrix} \xi \\ \eta \end{pmatrix}
\end{equation}
Please note, that two tuples $(\xi, \eta)$ and $(\alpha \xi, \alpha \eta)$ specify the same $z \in \mathbb{C}_\infty$ for any $\alpha \in \mathbb{C}\setminus\{0\}$. Therefore, they are called projective coordinates.

A different kind of crossed homomorphisms can then be defined, if $Q$ is a homogeneous, positive valued polynomial of degree $d$ in $\mathbb{C} \times \mathbb{C}$, i.e. a map
\begin{equation}
Q: \mathbb{C} \times \mathbb{C} \rightarrow \mathbb{R}^+ , (\xi, \eta) \mapsto Q(\xi, \eta)
\end{equation}
which satisfies $Q(\alpha \xi, \alpha \eta) = \alpha^d Q(\xi, \eta)$ for all $\alpha \in \mathbb{C}\setminus\{0\}$. It is then easy to show, that the associated map (where $\mathbb{C}_\infty$ is now coordinatized in projective coordinates)
\begin{equation}
c^{(Q)}: \mathrm{SO}^+(1,3) \rightarrow C^\infty(\mathbb{C}_\infty, \mathbb{R}^+), \Lambda \mapsto c_\Lambda^{(Q)} := \frac{Q(A_\Lambda ~\cdot~)}{Q(\cdot)},
\end{equation}
which should be understood as
\begin{equation}
c^{(Q)}_\Lambda(z = \xi/\eta) = \frac{ Q(A_\Lambda (\xi, \eta)^T)}{Q\left( \xi, \eta\right)}
\end{equation}
gives a well defined crossed homomorphism. Hence, the associated embedding
\begin{equation}
\mathrm{SO}^+(1,3) \hookrightarrow G, \Lambda \mapsto \left(A_\Lambda, \frac{Q(A_\Lambda ~\cdot~)}{Q(\cdot)} \right)
\end{equation}
defines then a Lorentz subgroup of $G$.

\subsection{Characterization of the subgroup of isometries}\label{5.3}
In the last subsection we saw, that $G$ comprises infinitely many Lorentz subgroups. In this section we will show, that the group of isometries $\mathrm{Iso}^+_p$ is one of those Lorentz subgroups, and that it is induced by the usual representation of the Lorentz group on $T_p \mathscr{M}$. Especially, we will present the crossed homomorphism to which the inclusion $\mathrm{Iso}^+_p \subset \mathrm{Con}^+_p$ is associated in section \ref{IC} and in section \ref{EC} we will explain, how the group $\mathrm{Iso}^+_p$ is induced by the action of the Lorentz group on $T_p\mathscr{M}$ associated with a local vielbein.
\subsubsection{Intrinsic characterization}\label{IC}
Recall, that we have defined in \eqref{fadef} a map $f^A$ associated to an $A \in \mathrm{PSL}(2, \mathbb{C})$ as the inverse of the conformal factor $K^A$ from \eqref{KF1}. By the use of projective coordinates, this map can be conveniently written as
\begin{equation}
f^A : \mathbb{C}_\infty \rightarrow \mathbb{C}_\infty, z = \xi/\eta \mapsto \frac{\begin{pmatrix} \bar{\xi} & \bar{\eta} \end{pmatrix} A^* A \begin{pmatrix} \xi \\ \eta \end{pmatrix}}{\begin{pmatrix}  \bar{\xi} & \bar{\eta} \end{pmatrix} \begin{pmatrix}  \xi \\ \eta  \end{pmatrix}}.
\end{equation}
This representation makes it then easy to show, that $f$ defines indeed a crossed homomorphism
\begin{equation}
f: \mathrm{PSL}(2, \mathbb{C}) \rightarrow C^\infty(\mathbb{C}_\infty, \mathbb{R}^+), A \mapsto f^A,
\end{equation}
by checking either
\begin{equation}
f^{A_1 A_2} = f^{A_1} \circ Z^{A_2} \cdot f^{A_2} 
\end{equation}
explicitely, or by realizing, that this map is, under the isomorphism $\mathrm{SO}^+(1,3) \cong \mathrm{PSL}(2, \mathbb{C})$,  an example of the second class of crossed homomorphisms as presented in section \ref{5.2}. The latter holds especially, because the map
\begin{equation}
Q: \mathbb{C} \times \mathbb{C} \rightarrow \mathbb{R}^+, (\xi, \eta) \mapsto \begin{pmatrix} \bar{\xi} &  \bar{\eta}\end{pmatrix}\begin{pmatrix} \xi \\ \eta \end{pmatrix} = \xi \bar{\xi} + \eta \bar{\eta}
\end{equation}
is a positive valued homogeneous polynomial of order $2$ on $\mathbb{C} \times \mathbb{C}$. By the isomorphism $\mathrm{SO}^+(1,3) \cong \mathrm{PSL}(2, \mathbb{C})$, this defines then a crossed homomorphism
\begin{equation}
f: \mathrm{SO}^+(1,3) \rightarrow C^\infty(\mathbb{C}_\infty, \mathbb{R}^+), \Lambda \mapsto f_\Lambda := f^{A_\Lambda}. \label{CHL}
\end{equation}
Recall now from section \ref{4.4}, that the isometry group $\mathrm{Iso}^+_p$ is explicitely given by
 \begin{equation}
\mathrm{Iso}^+_p = \left( \{ \Phi_p^{A} | A \in \mathrm{PSL}(2, \mathbb{C})\}, \circ \right)
\end{equation}
with
\begin{align}
\Phi_p^A : \mathbb{C}_\infty \times \mathbb{R}^+ \rightarrow \mathbb{C}_\infty \times \mathbb{R}^+, (z, \lambda) \mapsto (Z_A(z), f^A(z) \lambda).
\end{align}
Hence, 
\begin{equation}
\Phi_p^{A} = \Psi_p^{(A, f^A)}
\end{equation}
holds, where $\Psi_p^{(A, f^A)} \in \mathrm{Con}^+_p$ denotes the conformal automorphism associated with $(A, f^A) \in G$, as given by \eqref{PDF}. Hence,
\begin{equation}
\mathrm{Iso}^+_p \cong i_f(\mathrm{SO}^+(1,3) ) \subset G
\end{equation}
is satisfied, where $i_f$ is the embedding (\ref{emb11}) associated with the crossed homomorphism $f$, i.e.:
\begin{equation}
i_f: \mathrm{SO}^+(1,3) \hookrightarrow G, \Lambda \mapsto (A_\Lambda, f_\Lambda). \label{SOE}
\end{equation}
Consequently, $\mathrm{Iso}^+_p$ is isomorphic to the Lorentz subgroup of $G$ that is specified by the crossed homomorphism (\ref{CHL}).
\subsubsection{Extrinsic characterization}\label{EC}
We now want to show, how the group $\mathrm{Iso}^+_p$ arises equally in terms of the representation of the Lorentz group $\mathrm{SO}^+(1,3)$ on $T_p\mathscr{M}$ associated with a local vielbein. Let therefore $(U, \psi)$ be a local trivialization with $p \in U$ and $(E_\mu)$ be an associated vielbein. We then obtain an associated action of $\mathrm{SO}^+(1,3)$ on $T_p\mathscr{M}$ defined as
\begin{equation}
\Lambda v := \left( \Lambda^{\mu}_{~\nu} v^\nu \right) E_\mu
\end{equation}
for any vector $v = v^\mu E_\mu \in T_p\mathscr{M}$. Let now
\begin{equation}
\psi_p^+: L_p^+\mathscr{M} \rightarrow \mathbb{C}_\infty \times \mathbb{R}^+, v \mapsto \left(z_p^\psi(v), \lambda_p^\psi(v)\right)
\end{equation}
be the coordinate system in $\mathcal{B}_p$ that is induced by $(U, \psi)$. We then have by the results of appendix \ref{LMCO3} and especially by (\ref{tra3}-\ref{tra4}), that for any null vector $v \in L^+_p\mathscr{M}$
\begin{equation}
\psi_p^+(\Lambda v) = (A_\Lambda, f_\Lambda) \star \psi_p^+(v)
\end{equation}
holds, with $(A_\Lambda, f_\Lambda) \in G$, $f_\Lambda$ being defined by \eqref{CHL} and $\star$ being the group action \eqref{star}. By the results of section \ref{5.2}, this can then equally be written as
\begin{equation}
\psi_p^+(\Lambda v) = \Phi^{A_\Lambda} \circ \psi_p^+(v).
\end{equation}
Moreover we have for $\Lambda_1, \Lambda_2 \in \mathrm{SO}^+(1,3)$:
\begin{equation}
\psi_p^+(\Lambda_1 \Lambda_2 v) = \left((\Lambda_1, f_{\Lambda_1})\cdot (\Lambda_2, f_{\Lambda_2})\right) \star \psi_p^+(v).
\end{equation}
And by this we see, that the action of $\mathrm{SO}^+(1,3)$ on $T_p\mathscr{M}$ induces the group $\mathrm{Iso}^+_p$. This can be equally understood by observing, that for all $\Lambda \in \mathrm{SO}^+(1,3)$ and all $v, w \in T_p\mathscr{M}$
\begin{equation}
g(\Lambda v, \Lambda w) = g(v, w)
\end{equation}
must hold. Hence, the restriction of any Lorentz transformation $\Lambda \in \mathrm{SO}^+(1,3)$ to $L_p^+\mathscr{M}$ constitutes an isometriy for the induced metric $q$ on $L_p^+\mathscr{M}$ from \ref{S4}.
\subsection{On length gauges and Lorentz subgroups}\label{OLG}
In this subsection we want to gain a better intuition for the physical interpretation of the group $G$ and its Lorentz subgroups. To do so, we will now adopt a "passive" point of view and try to understand, which coordinate systems for $L_p^+\mathscr{M}$ are induced, if one composes conformal automorphisms $\Psi \in \mathrm{Con}^+_p$ with a coordinate system $\psi^+_p \in \mathcal{B}_p$. But first, we take one step back and ask in full generality, if and how one could define meaningful notions of length for null vectors in $L_p^+\mathscr{M}$.

On the one hand, one could be tempted to think, that no such notion exists, since the inner product $g_p$ is degenerate on $L^+_p\mathscr{M}$ and hence
\begin{equation}
\sqrt{g_p(v, v)} = 0
\end{equation}
holds. On the other hand, $L^+_p\mathscr{M}$ is a linear cone and thus it is still possible to define meaningful notions of length as homogeneous, positive definite maps
\begin{equation}
\lambda_p: L^+_p \mathscr{M} \rightarrow \mathbb{R}^+,\label{NOL}
\end{equation}
which mimick hence the properites of vector space norms in the present situation. We will call in the sequel such a map \eqref{NOL} a length gauge for $L^+_p\mathscr{M}$. 

Now consider first a map $\psi^+_p \in \mathcal{B}_p$ with associated vielbein $(E_\mu)$. This map can be written as (cf. \ref{rest})
\begin{equation}
\psi_p^+: L_p^+\mathscr{M} \rightarrow \mathbb{C}_\infty \times \mathbb{R}^+, v \mapsto \left(z_p^\psi(v), \lambda_p^\psi(v) \right) \label{lpps0}
\end{equation}
with 
\begin{equation}
\lambda_p^\psi(v^\mu E_\mu) = v^0 = |\vec{v}| = \sqrt{(v^1)^2 + (v^2)^2 + (v^3)^2}. \label{lpps}
\end{equation}
Here, $\lambda_p^\psi$ as given by \eqref{lpps} is obviously a length gauge for $L_p\mathscr{M}$ and hence, any map $\psi^+_p \in \mathcal{B}_p$ determines a length gauge for $L_p^+\mathscr{M}$ as the $3$-length (or equivalently as the $0$-component) of a null vector in the associated vielbein frame. Hence, the class of coordinates $\mathcal{B}_p$ (or equivalently the bundle atlas $\mathcal{B}$) singles out a class of length gauges for $L_p^+\mathscr{M}$, namely exactly those, that are associated to vielbein frames, as specified by \eqref{lpps}.

At this stage, at least the author feels a bit uncomfortable: The $3$-length (or equivalently the zero component $v^0$) associated to a vielbein frame is not a Lorentz invariant quantity. Why should it be hence a preferred notion of length for null vectors? For example, Penrose writes in \cite{Pen84}:
\begin{quote}
\textit{The extent of a null vector cannot be characterized in an invariant way by a number, nor can null vectors of different directions be compared with respect to extent. The ratio of the extents of null vectors of the same direction is meaningful, being just the ratio of the vectors.}
\end{quote}
Here, Penrose calls it "the extent of a null vector", what we call the $3$-length as given by the length gauge (\ref{lpps}). Hence, the length gauges \eqref{lpps} as singled out by the class of coordinates $\mathcal{B}_p$ don't seem to be meaningful quantities. Now, there are two strategies, how one could deal with this insight. Either one could discard length gauges completely, what leads to the usual "identification" of a future pointing light cone with the celestial sphere, together with the corresponding well known theory (cf. \cite{Pen84}).  But instead of doing so, one could adopt a different strategy: Instead of discarding length gauges completely, one could consider contrarily the set of \textit{all} possible length gauges as an invariant geometric structure associated with $L_p^+\mathscr{M}$ and analyse its properties. And we will see now, that this is exactly what we did in this paper. Especially we will understand, that the group $G$ describes all possible length gauges together with their transformation properties under Lorentz transformations. 

Therefore consider again a coordinate system $\psi^+_p \in \mathcal{B}_p$ as given by \eqref{lpps0}. Now let $(\mathbf{1}, Y) \in G$ and consider the map $\psi^{Y}_p := \Psi^{(\mathbf{1}, Y) } \circ \psi^+_p = (\mathbf{1}, Y)  \star \psi^+_p$. This map will be explicitely written as
\begin{align}
\psi_p^Y: L^+_p \mathscr{M} \rightarrow \mathbb{C}_\infty \times \mathbb{R}^+, v \mapsto \left(z^Y_p(v), \lambda^Y_p(v) \right)
\end{align}
and by the very definition of the group action $\star$ (or equivalently by the action of conformal automorphisms as presented in section \ref{4.2}) we have then:
\begin{align}
z^Y_p(v) &= z^\psi_p(v), \\
\lambda^Y_p(v) &= Y(z^\psi_p(v)) \lambda^\psi(v). \label{NOL2}
\end{align}
And hence, since $Y \in C^\infty(\mathbb{C}_\infty, \mathbb{R}^+)$, any length gauge is induced, if we act with group elements $g \in G$ on coordinate systems in $\mathcal{B}_p$. I.e. any length gauge \eqref{NOL} can be written as \eqref{NOL2} for a suitable $Y \in C^\infty(\mathbb{C}_\infty, \mathbb{R}^+)$. Consequently, group elements $g \in G$ of the form $g = (\mathbf{1}, Y)$ could be understood as pure length gauge transformations. In addition one can now enlarge the class of coordinates $\mathcal{B}_p$ to a larger class $\mathcal{R}_p$ which incorporates all those possible length gauges, i.e.:
\begin{equation}
\mathcal{R}_p = \{ (\mathbf{1}, Y) \star \psi^+_p | Y \in C^\infty(\mathbb{C}_\infty, \mathbb{R}^+) \text{~and~} \psi^+_p \in \mathcal{B}_p \}. \label{Rpd}
\end{equation}
Now, by the results of section \ref{IC}, we have, that for any two $\psi^+_{1,p}, \psi^+_{2,p} \in \mathcal{B}_p$ there exists exactly one $g \in \mathrm{Iso}^+_p \subset \mathrm{Con}^+_p$, s.th. $\psi^+_{1,p} = g \star\psi^+_{2,p}$ holds. Hence, we have
\begin{equation}
\mathcal{R}_p = G \star \psi^+_p
\end{equation}
for any $\psi^+_p \in \mathcal{B}_p$. And moreover, it follows then, that $G$ acts simply transitive on $\mathcal{R}_p$ and hence parametrizes all possible coordinate systems for $L_p^+\mathscr{M}$ that incorporate all admissible length gauges thereon. 

But how do the various Lorentz subgroups enter the game now? This is maybe the most subtle and interesting aspect of the present discussion. Therefore consider first again maps $\psi_p^+ \in \mathcal{B}_p$ as given by \eqref{lpps0}. As explained above, the associated length gauges $\lambda_p^\psi$ given by \eqref{lpps} are $3$-lengths induced by vielbein frames. And as such, there is an associated transformation law under Lorentz transformations, explicitely described by
\begin{equation}
\psi^+_p(\Lambda v ) = (A_\Lambda, f_\Lambda) \star \psi_p^+(v)
\end{equation}
as presented in section \ref{EC}. But if one enlarges the class of coordinates, such that all length gauges are allowed, i.e. if one makes a transition from $\mathcal{B}_p$ to $\mathcal{R}_p$ as defined in \eqref{Rpd}, then there exists no single, preferred Lorentz transformation law anymore. Instead, there are infinitely many possible Lorentz transformation laws as described by the various Lorentz subgroups of $G$. Moreover, if we consider again the group antihomomorphism $\kappa$ from \eqref{AH1}, we see, that it is not possible to absorb any crossed homomorphism other than the trivial one in the group composition law. I.e. we can't define
\begin{equation}
\kappa_A: C^\infty(\mathbb{C}_\infty, \mathbb{R}^+) \rightarrow C^\infty(\mathbb{C}_\infty, \mathbb{R}^+) , Y \mapsto  c_A \cdot Y \circ Z^A 
\end{equation}
for $c_A \neq 1$, since otherwise $\kappa_A(1) = c_A \neq 1$ and hence, $G$ would not constitute a group. This means, that from a group theoretic perspective, no non-trivial Lorentz subgroup of $G$ could be preferred in terms of an alternative multiplication law for $G$. Especially we see, that the crossed homomorphism $A \mapsto f_A$ seems completely arbitrary from this perspective. Conversely, any Lorentz subgroup of $G$ singles out a subclass of length gauges and a corresponding Lorentz transformation law, which is again unsatisfactory in the light of the discussion above surrounding the quote of Penrose. Hence it seems, that no Lorentz subgroup of $G$ and especially not its subgroup of isometries is intrinsically preferred. By this argumentation, $G$ could be considered as a more natural automorphism group for $L_p^+\mathscr{M}$ than any of its Lorentz subgroups.
\section{Comparison with the original BMS analysis}\label{6.1}
In this subsection, we want to compare the original BMS analyis at null infinity with our microscopic analogue. Especially, we will compare our methodology with the original BMS methodology in section \ref{6.12} and in section \ref{6.13} we will compare the structure of $G$ with the structure of the original BMS group. The similarities between those situations will then justify, why we call $G$ a microsocpic analogue of the BMS group, as will be summarized in section \ref{ConcBMS}.
\subsection{Methodology}\label{6.12}
We want to review concisely the original BMS analysis of asymptotically flat spacetimes in a modern language. Our main sources for this are \cite{Ger77, Ash14, Ash18}. To understand the original BMS analysis of asymptotic symmetries, it is instructive (cf. \cite{Ash14, Ash18}), to perform a conformal compactification (cf. \cite{Pen65}) of the asymptotically flat spacetime under consideration. Then null infinity $\mathcal{I}$ becomes a $3$-dimensional submanifold of Einstein's static universe, which can be regarded as a part of the topological boundary of the considered "physical" spacetime and as such, it inherits several universal geometric structures from the physical spacetime. Those structures (cf. \cite{Ash14}) consist out of a degenerate metric $q_{ab}$ with signature $(0, +, +)$ and a complete vector field $n$ on $\mathcal{I}$, which is defined in terms of the conformal factor $\Omega$ that was used for the conformal compactification:
\begin{equation}
n^\mu = \nabla^\mu \Omega
\end{equation} 
Here, $\nabla$ denotes the Levi-Civita connection on the compactified spacetime. In addition, there is a remaining freedom for admissible conformal transformations of the physical spacetime given by smooth redefinitions of the conformal factor $\Omega$ of the form
\begin{equation}
\Omega'= \omega \Omega.
\end{equation}
Here, $\omega$ is a smooth function on the compactified spacetime (including its boundary) that is nowhere vanishing on $\mathcal{I}$. One can restrict this "gauge freedom" further by allowing solely conformal factors, that satisfy $\nabla_\mu n^\mu = 0$. One can then show (cf. \cite{Ash14, Ash18}) that smooth factors $\omega$ preserve this condition if and only if their Lie derivative along $n$ vanishes, i.e. $\mathcal{L}_n \omega = 0$, under restriction to $\mathcal{I}$. Under such "gauge transformations", the universal structure $(q_{\mu \nu}, n^\mu)$ transforms as follows:
\begin{align}
q'_{\mu \nu} = \omega^2 q_{\mu \nu} & & n'^\mu = \omega^{-1} n^\mu \label{contra}
\end{align}
One then calls two tuples $(q_{\mu \nu}, n^\mu)$ and $(q'_{\mu \nu}, n'^\mu)$ equivalent, if they are related to each other by a conformal transformation of the type (\ref{contra}). The BMS group can then be understood as the group of all transformations on $I$, which preserve this universal structure, i.e. map tuples $(q_{\mu \nu}, n^\mu)$ to equivalent tuples.

The relation to our analysis becomes apparent, if one realizes, that all objects appearing in the original BMS analysis have direct microscopic analogues in our analysis. Consider first asymptotic flatness. In the case of the original BMS framework, the property of asymptotic flatness was exactly described in terms of decay conditions for the metric along null rays towards infinity (cf. \cite{Mad16}). As explained in sections \ref{1} and \ref{2.2}, our analogue to asymptotic flatness is given by Einstein's equivalence principle. As sketched in footnote \ref{fn1}, Einstein's equivalence principle demands flatness in an infinitesimal limit, which can be quantified in terms of Riemann normal coordinates: Around any point $p \in \mathscr{M}$ there exists coordinate patch $(x^\mu)$ around $p$ (i.e. $x^\mu(p) = 0)$ in which the metric assumes the form (cf. \cite{Bre98, Ste12})
\begin{equation}
g_{\mu \nu}(x) = \eta_{\mu \nu} - \frac{1}{3} R_{\mu \alpha \nu \beta}(0) x^\alpha x^\beta + \mathcal{O}(|x|^3),
\end{equation} 
where $R_{\mu \alpha \nu \beta}$ is a coordinate expression for the Riemann tensor. In this formulation, Einstein's equivalence principle is quantified in terms of a microscopic asymptotic decay condition and hence resembles the role of asymptotic flatness in the BMS framework. Consider now null infinity: It is a natural macroscopic null surface associated with any asymptotically flat spacetime and can be represented as the union of a pointed past and a pointed future light cone in Einstein's static universe. Moreover, it is diffeomorphic to $\mathbb{C}_\infty \times \mathbb{R}$. Analogously, the past and future tangent light cones $L_p^+\mathscr{M}$ are natural microscopic null surfaces associated with any spacetime satisfying Einstein's equivalence principle and are diffeomorphic to $\mathbb{C}_\infty \times \mathbb{R}^+$. In the BMS analysis, the geometry of the bulk spacetime induces a degenerate metric $q$ and a complete null vector field $n^\mu$ on null infinity. In our situation, the geometry of $T_p\mathscr{M}$ together with the psuedo-Riemannian metric $g$ induces also a degenerate metric as given by \ref{S4} on $L_p^+ \mathscr{M}$. The analogue to the complete null vector field $n^\mu$ is then given by the linear cone structure of $L_p^+\mathscr{M}$. The microscopic analogue to the gauge condition $\mathcal{L}_n \omega$ in the original BMS-analysis is the requirement, that conformal transformations should preserve the compatibility of the induced metric with the linear cone structure, as depicted at the end of section \ref{4.3}. The group $\mathrm{Con}^+_p$ is in our situation then the group which preserves the linear cone structure and the conformal structure on $L_p^+\mathscr{M}$ up to conformal equivalence, analogously as the BMS group preserves the universal structure in the sense depicted above.

\subsection{Group structure}\label{6.13}
In this section we will compare the structure of $G$ with the structure of the original BMS group. Thereby we will see, that some subtile differences appear, although the structure of $G$ and the structure of the original BMS group are still very similar. Our main source regarding the structure of the original BMS group will be the modern review \cite{Esp18}. Classic sources on this topic are \cite{Sa62, Can66, Mcc72}. Null infinity can be coordinatized in Bondi coordinates by $(z, u) \in \mathbb{C}_\infty \times \mathbb{R}$ (cf. \cite{Obl15}) and in \cite{Esp18} a general BMS transformation thereon is given by (cf. Formulas 5.15a and 5.15b of \cite{Esp18})
\begin{align}
z \rightarrow z' &= \frac{az + b}{cz+d} \label{TF1} \\
u \rightarrow u'&= K(z, \bar{z}) \left[ u + \alpha(z, \bar{z}) \right] \label{TF2}
\end{align}
where $K$ is the conformal factor given by
\begin{equation}
K(z, \bar{z}) = \frac{1+z \bar{z}}{(a z + b) (\bar{a} \bar{z} + \bar{b}) + (cz + d) (\bar{c} \bar{z} + \bar{d})}, \label{KoF}
\end{equation}
and $\alpha \in \mathbb{C}^\infty(\mathbb{C}_\infty, \mathbb{R}^+)$ is a supertranslation. Observe now, that the formulas (\ref{TF1} - \ref{TF2}) are very similar to  the coordinate expression of a transformation of $L_p^+\mathscr{M}$ that corresponds to an iterative application of a generic element $\Phi^{A}_p \in \mathrm{Iso}^+_p$ and an element $\Psi^{(\mathbf{1}, Y)}_p \in \mathrm{Con}^+_p$, since
\begin{equation}
\Psi^{(\mathbf{1}, Y)}_p \circ \Phi^{A}_p(z, \lambda) = \left(Z^A(z), Y(z) f^A(z) \lambda\right).
\end{equation}
holds, what means, that $(z, \lambda)$ transforms under $\Psi^{(\mathbf{1}, Y)}_p \circ \Phi^{A}_p \in \mathrm{Con}^+_p$ as
\begin{align}
z \rightarrow z' &= \frac{az + b}{cz + d} \label{TF3}\\
\lambda \rightarrow \lambda' &= K(z, \bar{z})^{-1} Y(z) \lambda \label{TF4}
\end{align}
with the conformal factor $K$ given by (\ref{KoF}). The important differences between (\ref{TF3} - \ref{TF4}) and (\ref{TF1} - \ref{TF2}) are, that our analogues of supertranslations do not act by addition, but by multiplication, and that our analogue of the conformal factor, given by the crossed homomorphism $f_A$, is exactly the inverse of $K$. The former can be easily understood, since automorphisms of the form $\Psi^{(\mathbf{1}, Y)}$ are multiplicative "superrescalings" of null vectors. The occurence of the inverse conformal factor in \eqref{TF4} can be understood in the present situation, if one derives the Lorentz-Möbius correspondence not as we will do it in appendix \ref{LMCO3}, but in terms of Bondi coordinates. Then one obtains, that the advanced coordinate $u = t + r$ rescales under Lorentz transformations exactly under the inverse prefactor as the radial coordinate $r$. This is for example presented in sections 4.2.1 and 4.2.2 of \cite{Obl15}.

In all articles, which we have mentioned above, the structure of the original BMS group is described as a semidirect product of the Lorentz group $\mathrm{SO}^+(1,3)$ with the group of supertranslations $C^\infty(\mathbb{C}_\infty, \mathbb{R})$. In this sense our group $G$ resembles the structure of the original BMS group. Especially in \cite{Esp18}, also the original BMS group is constructed as a right semidirect product, which resembles our construction of $G$ as performed in section \ref{5.1}. But if one digs deeper in the mentioned articles, an important difference will appear: In \cite{Esp18, Ger71}, the group action which is utilized for the definition of the semidirect product includes the conformal factor $K$. In particular, for a supertranslation $\alpha \in C^\infty(\mathbb{C}_\infty, \mathbb{R})$ and a Lorentz group element $\Lambda \in \mathrm{SO}^+(1,3)$, the group action is defined as (cf. eq. 6.14a-b of \cite{Esp18}):
\begin{equation}
\sigma_\Lambda(\alpha) = K^{-1} \cdot \alpha \circ Z_\Lambda. \label{Espa}
\end{equation}
Adapted to our situation, the corresponding modification of the group action $\kappa$ specified by (\ref{AH1} - \ref{AH2}) would be 
 \begin{equation}
 \kappa: \mathrm{PSL}(2, \mathbb{C}) \rightarrow \mathrm{Aut}(C^\infty(\mathbb{C}_\infty, \mathbb{R}^+)), A \mapsto \kappa_A \label{AH1b}
 \end{equation}
 with:
 \begin{equation}
 \kappa_A: Y \mapsto f_A \cdot Y \circ Z^A \label{AH2b}
 \end{equation}
 But this would not be appropriate in our situation. If we had adopted the group action (\ref{AH1b} - \ref{AH2b}), we would have had especially
 \begin{equation}
 \kappa_A(1) = f_A \neq 1.
 \end{equation}
Hence $\kappa_A$ would not constitute an automorphism of $C^\infty(\mathbb{C}_\infty, \mathbb{R}^+)$, since it would not preserve the unit element. Hence $G$ would in this case not constitute a proper group (cf. also the discussion at the end of section \ref{OLG}). Therefore we think, that $f_A$ must be excluded from the definition of the semidirect product in the present scenario. In the case of the original BMS group, this ambiguity does not appear, since it is additive. Hence its unit element is given by $0$ and fortunately the action (\ref{Espa}) satisfies $\sigma_\Lambda(0) = 0$. \\

\subsection{Conclusion}\label{ConcBMS}

We think, that the present discussion shows, that the original BMS group and the group $G$ are very similar: Both appear as conformal automorphism groups of natural null surfaces endowed with similar universal geometric structures, as depicted in section \ref{6.12}, and both can be written as right semidirect products of the Lorentz group (or equivalently the Möbius group) with a group of smooth functions on the Möbius sphere, as depicted in section \ref{6.13}.

\section{The automorphism groups as gauge groups for the light cone bundle}\label{GG}

In this section we want to sketch concisely, how the microscopic BMS-like group $G$, as well as its subgroup of isometries, constitute gauge groups for the bundle of future pointing null vectors. Therefore observe first, that the results from section \ref{2.2} generalize in a straightforward way to the full light cone bundle. I.e., the bundle $L^+\mathscr{M}$ is equipped with the following universal structures:
\begin{enumerate}[(F1)]
\item $F = \mathbb{C}_\infty \times \mathbb{R}^+$ is the typical fiber of the fiber bundle $L^+\mathscr{M}$, i.e. $L_p^+\mathscr{M} \cong F$. \label{F1}
\item Each fiber $L_p^+\mathscr{M}$ is a linear cone. \label{F2}
\item There is a family $(U_i, z_i)_{i \in I}$ of surjective maps
\begin{equation}
z_i: L^+U_i \rightarrow U_i \times \mathbb{C}_\infty
\end{equation}
whose transition functions are well defined and valued in the Möbius group. Here $(U_i)_{i \in I}$ is an open cover for $\mathscr{M}$. \label{F3}
\item There exists a degenerate metric $q_p$ on any fiber $L_p^+\mathscr{M}$. Moreover, the map
\begin{equation}
p \in \mathscr{M} \mapsto q_p
\end{equation}
is smooth. \label{F4}
\end{enumerate}

This allows then, as common in fiber bundle theory (cf. \cite{Ste99, Bau09, Ham17}), the definition of adapted bundle atlases that preserve those structures in a certain sense. The matching conditions of overlapping charts are then described in terms of transition functions that are valued in a structure group and analogously to the situation in section \ref{4}, the question thereby will be, how \ref{F4} should be interpreted: As a fixed degenerate Riemannian metric on $L_p^+\mathscr{M}$ or as a a representative of a conformal structure thereon? The former will yield a $\mathrm{SO}^+(1,3)$-structure for the lightcone bundle, while the latter gives rise to a $G$-structure.

Consider first the case, where $q_p$ is interpreted as a fixed degenerate Riemannian metric on any generic fiber $L_p^+\mathscr{M}$. Then the bundle atlas $\mathcal{B}$ as constructed in section \ref{2.1} is made up of smooth local trivializations $(U, \varphi)$ given by maps
\begin{equation}
\varphi: L^+ U \rightarrow U \times F
\end{equation}
which preserve the cone structure and have the property, that the metric $q_p$ is in each chart represented by
\begin{equation}
ds^2 = 2 \lambda^2 \frac{dz d\bar{z}}{(1 + z \bar{z})^2} \label{Mend}
\end{equation}
and whose transition functions lie, by direct generalization of the results of sections \ref{4.2} and \ref{EC}, in $\mathrm{Iso}^+_p \cong \mathrm{SO}^+(1,3)$. More explicitely, the latter means, that for two local trivializations $(U_1, \varphi_1), (U_2, \varphi_2) \in \mathcal{B}$ with $U_1 \cap U_2 \neq \emptyset$ the transition function
\begin{equation}
\varphi_2 \circ \varphi_1^{-1}: (U_2 \cap U_1)  \times F \rightarrow (U_2 \cap U_1) \times F
\end{equation}
can be written as
\begin{equation}
\varphi_2 \circ \varphi_1^{-1}(p, (z, \lambda) ) = (p, i_f(\Lambda_p) \star (z, \lambda) ),
\end{equation}
where 
\begin{equation}
\Lambda: U_1 \cap U_2 \rightarrow \mathrm{SO}^+(1,3), p \mapsto \Lambda_p
\end{equation}
is a local Lorentz transformation, $i_f$ is the embedding
\begin{equation}
i_f: \mathrm{SO}^+(1,3) \hookrightarrow G
\end{equation}
defined in (\ref{SOE}) and $\star$ is the group action (\ref{star}). In this sense, $\mathcal{B}$ constitutes a $\mathrm{SO}^+(1,3)$-structure for $L^+\mathscr{M}$, where the Lorentz group is non-trivially represented in terms of the crossed homomorphism $f$. 

We now interpret $q_p$ as a representative of a conformal equivalence class of metrics on a generic fiber $L_p^+\mathscr{M}$. Then the structures \ref{F1} - \ref{F4} induce an adapted bundle atlas, whose smooth trivializations $(U, \varphi)$ given by
\begin{equation}
\varphi: L^+U \rightarrow U \times F
\end{equation}
have the property, that the metric $q_p$ is in each chart $(U, \varphi)$ represented by
\begin{equation}
ds^2 = 2 \lambda^2 \Omega_p^\varphi(z)^2 \frac{dz d\bar{z}}{(1+z \bar{z})^2} \label{CF0}
\end{equation}
for an associated smooth conformal factor
\begin{equation}
\Omega^\varphi: U \times \mathbb{C}_\infty \rightarrow \mathbb{R}^+, (p,z) \mapsto \Omega_p^\varphi(z), \label{CF}
\end{equation}
and whose transition functions lie hence in $G$. The latter means, that for two such local trivializations $(U_1, \varphi_1), (U_2, \varphi_2)$ with $U_1 \cap U_2 \neq \emptyset$ the transition function
\begin{equation}
\varphi_2 \circ \varphi_1^{-1}: (U_2 \cap U_1)  \times F \rightarrow (U_2 \cap U_1) \times F
\end{equation}
can be written as
\begin{equation}
\varphi_2 \circ \varphi_1^{-1}(p, (z, \lambda) ) = (p, g_p \star (z, \lambda) )
\end{equation}
for 
\begin{equation}
g: U_1 \cap U_2 \rightarrow G, p \mapsto g_p
\end{equation}
being a local smooth\footnote{Of course, one has to specify, which notion of smoothness is meant exactly, since $G$ is infinite dimensional. But since we just want to sketch the ideas in the present discussion, we won't bother with this question here.} $G$-gauge transformation. In this sense, those local trivializations form a $G$-structure for $L^+\mathscr{M}$, which we call $\mathcal{R}$. Please note, that such an $G$-structure $\mathcal{R}$ indeed exists and can be constructed explicitely in terms of the atlas $\mathcal{B}$, by applying smooth $G$-valued maps locally on coordinate systems in $\mathcal{B}$. I.e. we define
\begin{equation}
\mathcal{R} = \left\{ (U, g \star \psi^+) \middle|  (U, \psi^+) \in \mathcal{B} \text{~and~} g \in C^\infty(U, G) \right\}
\end{equation}
where $g \star \psi^+$ is defined as:
\begin{equation}
g \star \psi^+: L^+U \rightarrow U \times F, v \in L^+_p\mathscr{M} \mapsto \left( p, g_p \star \psi^+_p(v)\right)
\end{equation}
By this, it follows directly, that the restrictions of maps in $\mathcal{R}$ to $T_p\mathscr{M}$ are given by $\mathcal{R}_p$ as constructed in section \ref{OLG}.

In this sense, we have identified two extremal natural gauge groups for the bundle $L^+\mathscr{M}$: The Lorentz group $\mathrm{SO}^+(1,3)$ forms a kind of minimal gauge group, while the microscopic BMS group $G$ constitutes some sort of maximal gauge group. By the argumentation of section \ref{OLG}, the former can be understood as a gauge group, which carries already information on the embedding $L^+\mathscr{M} \hookrightarrow T\mathscr{M}$, while the group $G$ seems to be preferred by the intrinsic geometry of $L^+\mathscr{M}$. Nevertheless, there should be some hard criterion, by which one can answer the question, which of those gauge groups is the "correct one". We will comment on this again in section \ref{PI}. Please note in addition, that, by the results of this section, we have now all necessary ingredients for a $\mathrm{SO}^+(1,3)$- and a $G$-gauge theory on $L^+\mathscr{M}$. I.e. one could define associated principal fiber bundles and analyze connections thereon.

Finally we would like to note, that there are also groups $H$ which satisfy
\begin{equation}
\mathrm{SO}^+(1,3) \subsetneq H \subsetneq G
\end{equation}
and could be also understood as possible gauge groups for $L^+\mathscr{M}$. For example, one could restrict the class of the allowed conformal factors $\Omega^\varphi_p(z)$ from equation (\ref{CF0}). Also, any of the Lorentz subgroups of $G$ should induce a corresponding bundle atlas. A more rigorous mathematical investigation of those structures as well as their geometric interpretations would be desirable, and we will comment on this question again in section \ref{OOQ}.

\section{Discussion}\label{Conc}
In this section we want to conclude this article. Especially we will summarize our findings in section \ref{SUM}. In section \ref{PI} we will discuss possible implications of the present investigation, that are motivated by the various applications of the original BMS group and are associated with general relativity and the UV-structure of gauge theories. In section \ref{OOQ} we will comment on other open questions as well as on the relation to the existing literature.
\subsection{Summary}\label{SUM}
In this article we have performed a thorough analysis of the universal structures that are induced on the infinitesimal tangent light cones of a generic spacetime obeying Einstein's equivalence principle. Thereby, we obtained as a main result, that those structures single out two natural microscopic symmetry groups, that arise as automorphism groups: A non-trivially represented Lorentz group as their isometry group and a group $G$ as their conformal automorphism group. We investigated the mathematical structure of the group $G$ and showed, that it can be described in terms of a right semidirect product of the Lorentz group (or equivalently the Möbius group) with a group of smooth, positive valued functions on the Riemann sphere. We further showed, that $G$ contains infinitely many Lorentz subgroups which are parametrized in terms of crossed homomorphisms. We have demonstrated, how the isometry group arises as a non-canonical subgroup of $G$ and argued, that no Lorentz subgroup seems to be intrinsically preferred from a geometric and a group theoretic perspective. Especially, we realized thereby, that $G$ encodes all possible length gauge choices for null vectors and that any Lorentz subgroup corresponds to a subclass of such length gauge choices together with an an associated Lorentz transformation law. We also compared our methodology and results with the classic BMS analysis, and justified by this, that $G$ can be called a microscopic analogue of the BMS group. Finally, we have sketched, how $G$ and the isometry subgroup could constitute gauge groups for the bundle of null directions. By this we have identified a geometric structure which exists on the bulk of any spacetime obeying Einstein's equivalence principle and which is associated with a BMS-like group. This implies especially, that BMS-like groups do not only describe macroscopic asymptotic symmetries in general relativity, but also constitute a fundamental and, to the best of our knowledge, unknown microscopic symmetry of Lorentzian geometry/ This symmetry encodes in an invariant way, how null vectors interfere with Lorentz transformations. In addition we would like to mention three results of this article, that lie not in the mainline of argumentation, but are still worthwhile to be mentioned explicitly:
\begin{itemize}
\item We gave a convenient representation for the rescalings of null vectors under Lorentz transformations, cf. \eqref{resc11} or section \ref{EC}. This transformation law was derived by a reinterpretation of the implicit definition of the inverse stereographic projection \eqref{RISP}, which is convenient for the calculation of null vector rescalings, cf. appendix \ref{LMCO3}. Although it was of course realized all over literature, that null vectors transform under Lorentz transformation not only by a change of their direction, but also by a rescaling of their length (cf. e.g. \cite{Pen84, Mcc72, Sa62}), the representation \eqref{resc11} as well as its derivation in appendix \ref{LMCO3} are to the best of our knowledge new.
\item We have shown in section \ref{2.2}, that the square root of the negative standard Minkowski inner product constitutes a kind of distance function on the light cone of Minkowski vector space. Moreover we have shown, that this distance function is related to the chordal distance on the Riemann sphere. Although this result seems very elementary, it is, to the best of our knowledge, not present in the existent literature. Nevertheless, a related reasoning was performed in \cite{Pen74}.
\item The occurence of infinitely many Lorentz subgroups of the conformal automorphism group can be equally understood by the statement, that there seems to be no distinguished Lorentz transformation law for length gauges anymore, if one considers all possible length gauges for null vectors, cf. the end of section \ref{OLG}. I.e. if one enlarges the class of coordinate systems as described at the end of section \ref{OLG}, then the subgroup of isometries together with its associated Lorentz transformation law seem to loose their preferred role, and infinitely many Lorentz transformation laws emerge. 
\end{itemize}
Finally, we want to remark, that it is in the view of the author interesting, that the group $G$ carries no obvious canonical structure that singles out a non-trivial Lorentz subgroup. Usually, structure groups in fiber bundle theory are some kind of "group theoretic mirror" of the geometry under consideration, since they encode geometric properties of the fiber bundle (and its base manifold) in terms of group theoretic properties (cf. \cite{Bau09, Ste99}). In the present scenario, the metric \eqref{Mend} is a distinguished geometric object on $L^+_p\mathscr{M}$, since it has constant positive curvature. But as discussed at the end of section \ref{OLG}, the associated Lorentz subgroup of isometries seems not to be preferred from a group theoretic perspective. This could be interpreted as a hint, that the structure of $G$ favours a conformal interpretation of the induced metrics on $L^+_p \mathscr{M}$.

\subsection{On possible implications}\label{PI}
As said in the introduction, the original BMS group found various applications in different branches of modern gravitational and high energy physics. Motivated by this success and by the similarity of the microscopic BMS-like group $G$ presented in this article as compared to the original BMS group, we ask the question, if the group $G$ together with its associated geometric structures could have similar applications. Therefore we will sketch three directions of research, that are motivated by the findings of the present article and by applications of the original BMS group. The scenarios depicted in this subsection will be investigated in a subsequent series of publications.
\paragraph{A bulk description for gravitational waves?}
The original BMS analysis of universal structures and associated symmetries at null infinity was a conerstone of the proof, that gravitational waves indeed exist in general relativity (cf. \cite{Mad16}). The reason for this is, that the description of gravitational radiative degrees of freedom simplifies drastically, if one formulates them in terms of induced higher order quantities on null infinity (cf. \cite{Ash81, Ash14, Ash18}). Given the similarity of the group $G$ with the original BMS group and the similarity of the universal structures on $L_p^+\mathscr{M}$ with the universal structures on null infinity, one could ask, if an analysis of $G$-connections on the light cone bundle $L^+\mathscr{M}$ could also lead to a simplified description of radiative degrees of freedom on the bulk. I.e., in a more colloquial language, if infinitesimal light cones could serve as convenient probes for gravitational waves on the bulk. This could be indeed meaningful, since the equivalence classes of connections which appear at null infinity (cf. \cite{Ash81, Ash14, Ash18}) could correspond in our scenario to a single gauge equivalence class of $G$-connections on $L^+\mathscr{M}$ as induced by several inequivalent Levi-Civita connections on $T\mathscr{M}$. Consequently, also quantities like the Bondi news tensor or the leading order Weyl tensor (for both cf. \cite{Ash14}) could have analoga for $G$-connections. Finally, the tangent bundle of null infinity could be understood as an arena for boundary values of the bulk light cone bundle. By this, the description of radiative degrees of freedom on null infinity should correspond directly to boundary values for $G$-connections, which could define soliton-like vacuums solutions. 

\paragraph{An "UV-Triangle"?}
Recently, a deep interconnection between the original BMS group, soft theorems in quantum gauge theory and memory effects in gravitational physics was discovered, going under the name of the IR-triangle (cf. \cite{Str18}). Given the structural similarity of the group $G$ and the BMS group, one could ask the question, if there could be a similar interrelation between $G$ and the UV-structure of gauge theories. This seems appealing: Due to the infinite dimensionality of $G$, there could exist infinitely many charges associated with $G$. Similarly, as the charges of the original BMS group imply soft theorems in terms of their Ward identities, one could ask, if identities associated with $G$ could constrain scattering amplitudes non-perturbatively in the deep UV. Moreover, the existence of distinct Lorentz subgroups of $G$ could be related to a microscopic gravitational memory effect: A bypassing gravitational wave could link two Lorentz subgroups of $G$ to each other, resembling the situation at null infinity, where gravitational radiation links distinct Minkowski vacua to each other, which are otherwise related by supertranslations (cf. \cite{Str16, Ash14, Ash18}). \\

\paragraph{An extension of the fundamental gauge group of gravity? }
Finally we would like to propose two speculative scenarios, in which $G$ (or maybe also a subgroup of $G$ that is strictly larger than the subgroup of isometries) could constitute a fundamental gauge group for a full theory of gravity. But before doing so, please note first, that $G$ could indeed already arise as a gauge group for a specific, very realistic sector of general relativity: In the situation, where all test particles are assumed to be massless, the tangent bundle can be safely replaced by the light cone bundle and especially all gravitational quantities should influence such test particles only in terms of their induced quantities on $L^+\mathscr{M}$. One could then analyse, if Einstein's equation (or equivalently the Einstein-Hilbert action) can be reexpressed entirely in terms of induced quantities on $L^+\mathscr{M}$. By doing so, one should especially understand, which gauge freedom is dictated by the action principle (or equivalently by its canonical formulation) for the induced connections on $L^+\mathscr{M}$: Is the gauge freedom described by $G$ or is it described by the subgroup of isometries? The discussion of section \ref{OLG} makes it plausible, that the gauge group is indeed enlarged to $G$ in this scenario, but of course, this has to be investigated. But however, a careful analysis of this situation should shed in any case some light on the the question regarding the "correct" gauge group for the light cone bundle, as raised in section \ref{GG}. Please note also, that this scenario is also closely connected to a hypothetical bulk description of gravitational waves as sketched above: Both scenarios aim towards a simplification of general relativity by considering solely its effects on massless test particles. 

After sketching this realistic scenario regarding a subsector of general relativity, we now want to propose the first of two speculative scenarios, in which $G$ could constitute a gauge group for a full theory of gravity. Therefore recall, that we have shown in section \ref{5.2}, that $G$ contains infinitely many Lorentz subgroups. By the argumentation of section \ref{OLG} one can realize in addition, that two distinct Lorentz subgroups are related by some kind of length gauge transformation. Moreover, all Lorentz subgroups are (trivially) isomorphic to each other and especially isomorphic to the subgroup of isometries. This finding resembles to some extent the situation in spontaneously broken gauge theories and hence one could ask, if a Higgs-like mechanism could break the gauge symmetry of a $G$-gauge theory to an arbitrary Lorentz subgroup. By this, general relativity could constitute a low-energy approximation of such a theory, where the Higgs-like field is near to its ground state. Interestingly, everything should move at the speed of light in situations, where the Higgs-like field is not in its ground state. Please note in addition, that the subgroup of isometries is, although not intrinsically preferred, induced by the standard representation of the Lorentz group on $\mathbb{R}^4$. By this, the subgroup of isometries seems natural from the perspective of any spontaneously choosen Lorentz subgroup of $G$, although it seems unnatural if one considers $G$ as a fundamental gauge group.

For the second, more speculative scenario realize, that the group $G$ is really in some sense a conformal group. Hence, one could ask, how a conformal field theoy for $G$ would look like and if some sort of holographic principle could hold for $G$. If so, this holographic principle would be in some kind special, since it would not constitute a classic bulk-to-boundary correspondence, but merely a kind of "gauged holography", where quantities associated with "microscopic bulks" (as represented by the tangent spaces $T_p\mathscr{M}$) would be encoded on "microscopic boundaries" (as represented by tangent light cones $L^+_p\mathscr{M}$).  This idea seems appealing, since it would lie in between a pure boundary description and a pure bulk description. This is due to the fact, that in such a situation the "bulk theory" would be encoded microscopically in a $2$-dimensional way, since the typical fibers of the light cone bundle are $2$-dimensional geometric entities. Nevertheless, the base manifold and hence also the macroscopic effective geometry would still be $4$-dimensional. Please note in addition, that $G$ comprises infinitely many Lorentz subgroups, and hence it includes also infinitely many Lorentz transformation laws. In the light of this scenario, those subgroups could encode transformation properties of different physical microscopic bulk entities.

Finally note, that those scenarios seem to be plausible from different perspectives. There are various hints, that gravity behaves at fundamental scales in a $2$-dimensional way (cf. \cite{Car17}). In addition, the BKL-conjecture (cf. \cite{Bel70}) suggests, that gravity behaves at fundamental scales in an ultralocal way, which is also a property of ultrarelativistic field theories (cf. \cite{Dau98, Kla70}). Note, that those properties would be directly imprinted into any theory associated with $L^+\mathscr{M}$ and $G$: The light cone bundle is infinitesimally a $2$-dimensional space and hence the dimensional reduction at microscopic scales would be manifest in a $G$-gauge theory on the light cone bundle. Moreover, tangent light cones are obviously ultrarelativistic objects and the group $G$ should thus, analogously to the situation in \cite{Duv141, Duv142, Duv143}, be related to some ultrarelativistic symmetry group. Hence, also the ultralocal behaviour at microscopic scales should be manifest. This could be summarized in a picturesque way by saying, that a $G$-gauge theory on the light cone bundle should describe a situation, where fundamentally everything moves at the speed of light and where timelike causality relations are emergent by some yet to be invented mechanism. As argued above, this mechanism could be maybe a spontaneous symmetry breaking or a "gauged holographic principle".

\subsection{Other open questions and relation to the literature}\label{OOQ}
Finally we would like to comment on some other open questions and possible directions of research.
\paragraph{$G$ from the perspective of mathematical gauge theory:}
We have sketched in section \ref{GG}, how the group $G$ constitutes a gauge group for the light cone bundle. It could be an interesting question, if there exist manifolds, which admit a $G$-structure for a fiber bundle of linear tangent cones, but don't admit a pseudo-Riemannian metric. A further interesting question going in the same direction would be, how $G$-structures on the lightcone bundle interfere with spin structures on the manifold under consideration. Also, one could ask, as sketched concisely in section \ref{GG}, how subgroups of $G$ that are strictly larger than the isometry subgroup are related to geometric properties. Finally, it could be interesting to analyse, how the structures described in this article are related to global geometric and topological questions in pseudo-Riemannian geometry, by investigating the topology of the light cone bundle as well as its invariants (cf. \cite{Ste99}).
\paragraph{Towards a BMS geometry?}
It was shown in \cite{Duv141, Duv142}, that a certain conformal Carroll group of the macroscopic Minkowski light cone is given by the BMS-group. In contrast, we have shown in this article, that the conformal automorphism group of the infinitesimal light cone is given by $G$. The geometric relation of the original BMS group to the group $G$ resembles hence in some sense the relation of the Poincaré group to the Lorentz group: The BMS group acts on the macroscopic light cone, as the Poincaré group acts on the macroscopic Minkowski space, and the group $G$ acts on infinitesimal tangent light cones, as the Lorentz group acts on infinitesimal tangent Minkowski spaces. Now recall, that pseudo-Riemannian geometry can be written as a Cartan geometry (cf. \cite{Sha00}) based on the Poincaré and the Lorentz group. Although the structural interrelation of the group $G$ and the BMS group seem to forbid the formulation of a Cartan geometry based on those two groups, one could still ask the question, if above geometric picture could be interpreted in similar lines, giving rise to a kind of general BMS geometry in a, possibly extended, Cartan geometric framework.
\paragraph{The mathematical structure of $G$:}
As said in section \ref{5.2}, a general classification of crossed homomorphisms $c: \mathrm{SO}^+(1,3) \hookrightarrow G$ would be desirable. On the other hand, the existence of infinitely many Lorentz subgroups of $G$ suggests, that each of those subgroups could encode in some sense the microscopic transformation properties of some geometric or physical object. A better qualitative understanding of crossed homomorphisms associated with Lorentz subgroups could hence yield also a better understanding of the interpretation of the occuring Lorentz subgroups. Moreover, it would be very important in the context of the discussion of section \ref{OLG}, to understand, if the crossed homomorphism $f$ that defines the isometry subgroup is in some sense intrinsically preferred by the structure of $G$ or just extrinsically induced by the linear representation of the Lorentz group on $T_p\mathscr{M}$.
\paragraph{Relation to Carrollian geometry:}
In \cite{Duv141, Duv142} conformal Carroll groups were introduced and it was shown, that the BMS group is a conformal extension of the ultrarelativistic Carroll group (cf. \cite{Bac68}). Moreover, in \cite{Duv141} the notion of Carroll manifolds was introduced and it was shown, that the Minkowski light cone constitutes a Carroll manifold, whose conformal symmetry group is given by the BMS group. Those results are certainly related to the present analysis, but still distinct. First note, that, although the group $G$ and the BMS group are very similar,  they seem to be not isomorphic and have especially a different mathematical structure as demonstrated in section \ref{6.13}. But the most important difference is, that in our analysis infinitesimal tangent light cones were considered, whose interpretation and structure are, although similar, still different from their relatives in flat Minkowski spacetime. Although each tangent space is isomorphic to Minkowski \textit{vectorspace}, Minkowski \textit{spacetime} and a single tangent space are not equivalent as physical entities. Hence, their respective light cones also have different interpretations. This can be most easily seen by recapitulating from section \ref{GG}, that the BMS-like group $G$ can be considered as an eligible gauge group for a natural fiber bundle appearing on any spacetime. This result is new and does not follow from an analogous analysis in Minkowski spacetime.\\

\paragraph{Relation to residual gauge invariance in light cone gravity:} The conventional BMS algebra was recently found to describe a residual gauge freedom in 4-dimensional light cone gravity (\cite{Ana211, Ana212}). It is plausible, that this result could be related to our results. Nevertheless, the exact interrelation is unclear and should be investigated in the future.

\section*{Acknowledgements}
The author thanks Stefan Hofmann and Marc Schneider for inspiring discussions and for suggestions that increased the readability of the paper. Moreover, he acknowledges the Norbert Janssen foundation for financial support.
\newpage
\appendix
\section{The Lorentz-Möbius Correspondence as adapted to our situation}\label{LMCA}
In this section we want to explain the correspondence between the Lorentz and the Möbius group as adapted to our situation. Most of the results are adapted straightforwardly from literature, but some results are in the present form also new. To keep this section self-contained, we first review all basic facts on the Riemann sphere that are required for the understanding of this article, although they were partially already presented in the main body. This will be done in section \ref{LMCA1}. In section \ref{LMCA2} we will derive a useful interpretation for bundle trivializations in $\mathcal{B}$. In section \ref{LMCO3} we will then utilize this representation for the derivation of Lorentz transformation properties (\ref{mt11}, \ref{resc11}) of null vectors.
\subsection{The Riemann sphere and Möbius transformations}\label{LMCA1}
We review the basic theory of the Riemann sphere. Our main source for this section is given by \cite{Pen84}. The Riemann sphere is defined as the extended complex plane, i.e.
\begin{equation}
\mathbb{C}_\infty := \mathbb{C} \cup \{ \infty \}
\end{equation}
and is coordinatized by complex numbers $z \in \mathbb{C}_\infty$. But for the derivation of the Lorentz-Möbius correspondence a different set of coordinates will be more convenient. Those coordinates are given by tuples $(\xi, \eta) \in \mathbb{C}^2$ which are allowed to take any value other than $(0, 0)$ and will be called projective coordinates. A point $z \in \mathbb{C}_\infty$ on the Riemann sphere is then specified by the quotient
\begin{equation}
z = \xi/\eta.
\end{equation}
Observe, that two tuples $(\xi_1, \eta_1), (\xi_2, \eta_2)$ represent the same $z \in \mathbb{C}_\infty$ if and only if there is an $\alpha \in \mathbb{C} \setminus \{0 \}$ such that $(\xi_1, \eta_1) = (\alpha \xi_2, \alpha \eta_2)$ holds. This is the reason, why those coordinates are called projective coordinates (cf. \cite{Pen84}). 

The Riemann sphere $\mathbb{C}_\infty$ is diffeomorphic to the standard $2$-sphere $S^2$ in terms of the stereographic projection
\begin{equation}
\rho: S^2 \rightarrow \mathbb{C}_\infty, \hat{v} \mapsto \rho(\hat{v}) := \frac{\hat{v}^1 + i \hat{v}^2}{1 - \hat{v}^3},
\end{equation}
where we wrote a generic unit vector $\hat{v} \in S^2$ as $\hat{v} = (\hat{v}^1, \hat{v}^2, \hat{v}^3)$. The inverse of the stereographic projection will be denoted by $\hat{\epsilon} := \rho^{-1}$ and can be explicitely written as 
\begin{equation}
\hat{\epsilon}: \mathbb{C}_\infty \rightarrow S^2, z \mapsto \hat{\epsilon}(z) := \left(\hat{\epsilon}^1(z), \hat{\epsilon}^2(z), \hat{\epsilon}^3(z) \right)
\end{equation}
with (cf. \cite{Pen84}):
\begin{align}
\hat{\epsilon}^1\left( z = \xi/\eta\right) &= \frac{z + \bar{z}}{ \bar{z} + 1} = \frac{\xi \bar{\eta}+ \eta \bar{\xi}}{\xi \bar{\xi} + \eta \bar{\eta}} \label{ISP1} \\
\hat{\epsilon}^2\left( z = \xi/\eta \right) &= \frac{1}{i} \frac{z - \bar{z}}{z \bar{z} + 1} = \frac{1}{i} \frac{\xi \bar{\eta} - \eta \bar{\xi}}{\xi \bar{\xi} + \eta \bar{\eta}} \\
\hat{\epsilon}^3\left( z = \xi/\eta \right) &= \frac{z \bar{z} -1}{z \bar{z} + 1} = \frac{\xi \bar{\xi} - \eta \bar{\eta}}{\xi \bar{\xi} + \eta \bar{\eta}} \label{ISP3}
\end{align}
By introducing the Pauli matrices $(\sigma_\mu)_{\mu = 0, ..., 3}$ defined as
\begin{align}
\sigma_0 & = \begin{pmatrix} 1 & 0 \\ 0 & 1\end{pmatrix}, & 
\sigma_1 &= \begin{pmatrix} 0 & 1 \\ 1 & 0 \end{pmatrix}, \\
\sigma_2 &= \begin{pmatrix} 0 & i \\ -i & 0 \end{pmatrix}, &
\sigma_3&= \begin{pmatrix} 1 & 0 \\ 0 & -1 \end{pmatrix}.
\end{align}
and setting $\hat{\epsilon}^0(z) = 1$, we can write the relations \eqref{ISP1} - \eqref{ISP3} conveniently as
\begin{equation}
\hat{\epsilon}^\mu(z) \sigma_\mu = \frac{2}{\xi \bar{\xi} + \eta \bar{\eta}} \begin{pmatrix}\xi \\ \eta \end{pmatrix}\begin{pmatrix} \bar{\xi} & \bar{\eta} \end{pmatrix}  \label{RISP}
\end{equation}
for $z = \xi/\eta$ (cf. \cite{Pen84}).
We now want to introduce Möbius transformations as biholomorphic automorphisms of $\mathbb{C}_\infty$ that are given by (cf. \cite{Pen84, Jos06, Don11, Kno16})
\begin{equation}
Z: \mathbb{C}_\infty \rightarrow \mathbb{C}_\infty, z \mapsto \frac{az + b}{cz+d}
\end{equation}
for complex numbers $a,b,c,d \in \mathbb{C}$ satisfying $ad-bc = 1$. Strictly speaking, the requirement $ad - bc = 1$ is not necessary, but it is convenient, since it makes the structure of the Möbius group more lucid. Now, there exists a canonical surjective homomorphism (cf. \cite{Pen84}) between the group $\mathrm{SL}(2, \mathbb{C}) = \{ A \in \mathbb{C}^{2\times 2} | \mathrm{det}(A) = 1 \}$ and the Möbius group, which associates to a matrix
\begin{equation}
A = \begin{pmatrix} a & b \\ c& d\end{pmatrix} \in \mathrm{SL}(2, \mathbb{C})  \label{matrixm}
\end{equation}
the Möbius transformation 
\begin{equation}
Z^A(z) := \frac{az + b}{cz + d}.
\end{equation}
One can then see easily, that two matrices $A, B \in \mathrm{SL}(2, \mathbb{C})$ define the same Möbius transformation, i.e. $Z^A = Z^B$, if and only if $A  = \pm B$.  Having this in mind, writing matrices $A$ instead of Möbius transformations $Z^A$ is often more practical and one gets by this an isomorphism between the Möbius group and the group $\mathrm{PSL}(2, \mathbb{C}) := \mathrm{SL}(2, \mathbb{C}) \setminus \{ \pm 1\}$. Therefore, we denote the Möbius group often just by $\mathrm{PSL}(2, \mathbb{C})$. Moreover, we denote an equivalence class of matrices $[A]$ in $\mathrm{PSL}(2, \mathbb{C})$ just by one of their representatives, i.e. $[A] = A$ by slight abuse of notation. Observe in addition, that the action of a Möbius transformation can be easily expressed (by slight abuse of notation) in terms of projective coordinates as
\begin{equation}
Z^A\left(\xi/\eta\right)  = A \begin{pmatrix} \xi \\ \eta \end{pmatrix},
\end{equation}
which will be useful for the derivation of the Lorentz-Möbius correspondence.
Finally we want to introduce a distance function on $\mathbb{C}_\infty$ given by
\begin{equation}
d: \mathbb{C}_\infty \times \mathbb{C}_\infty \rightarrow [0, \infty), (z_1, z_2) \mapsto \frac{2 |z_1 - z_2| }{\sqrt{|z_1|^2 +1} \sqrt{|z_2|^2 + 1}}. \label{CDApp}
\end{equation}
This distance function makes $\mathbb{C}_\infty$ to a metric space and is the so called chordal distance (cf. \cite{Hil73}). It is induced by the corresponding euclidean distance of the associated points on $S^2$:
\begin{equation}
d(z_1, z_2) = \left| \hat{\epsilon}(z_1) - \hat{\epsilon}(z_2) \right| \label{CDApp2}
\end{equation}
The chordal distance is moreover related to a Riemannian metric on $\mathbb{C}_\infty$ that is explicietly given by:
\begin{equation}
ds^2 = \frac{4}{(1 + z \bar{z})^2} dz d\bar{z} \label{CMA}
\end{equation}
This metric is compatible with the natural conformal structure that $\mathbb{C}_\infty$ carries as a Riemann surface (cf. \cite{Jos06}). The pullback of the metric \ref{CMA} under a Möbius transformation $Z^A$ specified by a matrix \ref{matrixm} can be easily calculated as
\begin{equation}
ds^2 = K^A(z)^2 \frac{4}{(1+z \bar{z})^2} dz d\bar{z}
\end{equation}
where the conformal factor $K^A$ is given by:
\begin{align}
K^A(z) = \frac{1 + z \bar{z}}{(az + b) (\bar{a} \bar{z} + \bar{b}) + (cz + d) (\bar{c} \bar{z} + \bar{d})}
\end{align}
It can be written equivalently as
\begin{equation}
K^A(z) = \frac{1 + z \bar{z}}{1 + Z^A(z) \overline{Z^A(z)}} \frac{1}{(cz + d) (\bar{c} \bar{z} + \bar{d})}
\end{equation}
or by the utilization of projective coordinates as:
\begin{equation}
K^A(z) = \frac{\begin{pmatrix} \bar{\xi} & \bar{\eta} \end{pmatrix} \begin{pmatrix} \xi \\ \eta \end{pmatrix}}{\begin{pmatrix} \bar{\xi} & \bar{\eta} \end{pmatrix} A^* A \begin{pmatrix} \xi \\ \eta \end{pmatrix}}
\end{equation}
\subsection{Spin representation of light cone bundle trivializations}\label{LMCA2}
We now want to utilize the results of the last section, to derive a convenient representation for local trivializations $(U, \psi^+) \in \mathcal{B}$. This representation is an original construction of this article and is especially a reinterpretation of the relation \eqref{RISP}. It will then allow a very fast computation of null vector rescalings under Lorentz transformation and will hence be useful in the next subsection, where we finally present the Lorentz-Möbius correspondence as adapted to our situation. Therefore let $(U, \psi) \in \mathcal{A}$ be a local trivialization of $T\mathscr{M}$ with associated vielbein $(E_\mu)$ and let 
\begin{equation}
\psi^+: L^+U \rightarrow U \times (\mathbb{C}_\infty \times \mathbb{R}^+), v \in T_p\mathscr{M} \mapsto \left(p, (z_p^\psi(v), \lambda_p^\psi(v)\right) \label{btA}
\end{equation}
be the associated bundle trivialization of $L^+\mathscr{M}$ as defined in \eqref{AM1}. I.e. we have explicitely:
\begin{align}
z_p^\psi( v^\mu E_\mu) &= \rho(\hat{v}) = \frac{v^1 + i v^2}{v^0 - v^3} \\
\lambda_p^\psi(v^\mu E_\mu) &= |\vec{v}|
\end{align}
Let $p \in U$. As before, we denote the restriction of $\psi^+$ to $L_p^+\mathscr{M}$ by 
\begin{equation}
\psi^+_p: L_p^+\mathscr{M} \rightarrow L_p^+\mathscr{M}, v \mapsto (z^\psi_p(v), \lambda^\psi_p(v) ). \label{restA}
\end{equation}
We now want to generalize the relation \eqref{RISP}, such that we can use it as an implicit definition of such bundle trivializations. Therefore observe first, that by \eqref{RISP} any null vector $v \in L^+_p\mathscr{M}$ given in the vielbein frame $(E_\mu)$ by $v = v^\mu E_\mu$ can be written as
\begin{equation}
v^\mu \sigma_\mu = v^0 \frac{2}{\xi \bar{\xi} + \eta \bar{\eta}} \begin{pmatrix} \xi \\ \eta \end{pmatrix} \begin{pmatrix} \bar{\xi} & \bar{\eta} \end{pmatrix} \label{spin0}
\end{equation}
for a unique tuple $(z = \xi/\eta, v^0) \in \mathbb{C}_\infty \times \mathbb{R}^+$. This follows just by multiplication of \eqref{RISP} with $v^0$ and by $|\hat{\epsilon}(z)| = 1$. Note, that the mentioned projective freedom in the coordinates $(\xi, \eta)$ does not spoil the uniquenes of the representation (\ref{spin0}), due to the occurence of the factor $(\xi \bar{\xi} + \eta \bar{\eta})^{-1}$ on the right hand side of (\ref{spin0}). Equation \eqref{spin0} gives hence a correspondence between $L_p^+ \mathscr{M}$ and $\mathbb{C}_\infty \times \mathbb{R}^+$ associated to a vielbein frame $(E_\mu)$. In this formulation, one realizes, that \eqref{spin0} is an implicit definition of the map $\psi^+_p$ from \eqref{restA}. I.e. one could have defined the diffeomorphism \eqref{restA}, and hence also the bundle trivialization (\ref{btA}), equivalently by demanding, that
\begin{align}
v^\mu \sigma_\mu =&\lambda_p^\psi(v)  \frac{2}{\xi_p(v) \bar{\xi}_p(v) + \eta_p(v) \bar{\eta}_p(v)} \begin{pmatrix} \xi_p(v) \\ \eta_p(v) \end{pmatrix} \begin{pmatrix} \bar{\xi}_p(v) & \bar{\eta}_p(v) \end{pmatrix} \label{spin1}
\end{align}
should hold for all $v \in L^+_p \mathscr{M}$ with $v = v^\mu E_\mu$. Here, we have set $z_p^\psi(v) =: \xi_p(v)/\eta_p(v)$. This relation \eqref{spin1} will be called the spin representation of the local bundle trivialization \eqref{restA} and will be very useful in the next subsection. 
\subsection{The Lorentz-Möbius correspondence and associated null vector rescalings}\label{LMCO3}
We now present the well known correspondence between the Lorentz group $\mathrm{SO}^+(1,3)$ and the Möbius group $\mathrm{PSL}(2,\mathbb{C})$ as adapted to our situation. The ideas of this section come from \cite{Pen84} but are of course adapted to our situation. However, the utilization of the spin representions \eqref{RISP} and \eqref{spin1} for the derivation of null vector rescalings under Lorentz transformations is, to the best of our knowledge, an original development of this article. Therefore we fix again a $p \in \mathscr{M}$ and a vielbein frame $(E_\mu)$ associated to a local trivialization $(U, \psi) \in \mathcal{A}$ with $p \in U$. Let now $v \in L^+_p\mathscr{M}$ be a null vector with $ v = v^\mu E_\mu$. Then, as explained in the last subsection, the induced coordinate system
\begin{equation}
\psi_p^+: L^+_p \mathscr{M} \rightarrow \mathbb{C}_\infty \times \mathbb{R}^+ , v  \mapsto \left( z_p^\psi(v), \lambda_{p}^\psi(v)\right) 
\end{equation}
given by \eqref{restA} is implicitely defined by the relation \eqref{spin1}. Now forget first about the right hand side of \eqref{spin1}. The well known correspondence between the Lorentz- and the Möbius group states then (cf. \cite{Pen84, Obl15}), that there is $1$-to-$1$-correspondence between Lorentz transformations $\Lambda \in \mathrm{SO}^+(1,3)$ and matrices $A_\Lambda \in \mathrm{PSL}(2, \mathbb{C})$ such that
\begin{equation}
\left( \Lambda^{\mu}_{~\nu} v^\nu \sigma_\mu \right) = A_\Lambda \left(  v^\mu \sigma_\mu \right)A_\Lambda^* \label{corrl}
\end{equation}
holds for any $(v^\mu) \in \mathbb{R}^4$. The map $\Lambda \mapsto A_\Lambda$ constitutes then the isomorphism $\mathrm{SO}^+(1,3) \cong \mathrm{PSL}(2,\mathbb{C})$ (cf. \cite{Obl15, Pen84}).
By considering now again the right hand side of (\ref{spin1}), we are then able to derive, how an active local Lorentz transformation induces a transformation on $\mathbb{C}_\infty \times \mathbb{R}^+$. Therefore let $\Lambda \in \mathrm{SO}^+(1,3)$ be a Lorentz transformation and set $w = w^\mu E_\mu \in L^+_p\mathscr{M}$ with:
\begin{equation}
w^\mu =  \Lambda^{\mu}_{~\nu} v^\nu
\end{equation}
Set further $z_p^\psi(v) = \xi/\eta$. We then can write by (\ref{spin1}) and (\ref{corrl}):
\begin{equation}
 \left(  w^\mu \sigma_\mu \right) = \lambda_p^\psi(v) \frac{2}{\xi \bar{\xi} + \eta \bar{\eta}} A_\Lambda \begin{pmatrix} \xi \\ \eta \end{pmatrix} \begin{pmatrix} \bar{\xi} & \bar{\eta} \end{pmatrix} A^*_{\Lambda} \label{tw2}
\end{equation}
Set now
\begin{equation}
\begin{pmatrix}
\xi' \\ \eta'
\end{pmatrix} := 
A_\Lambda \begin{pmatrix} \xi \\ \eta \end{pmatrix}.
\end{equation}
Then, (\ref{tw2}) can be equally written as:
\begin{equation}
w^\mu \sigma_\mu = \left(  \lambda_p^\psi(v) \frac{\xi'\bar{\xi}'+ \eta'\bar{\eta}'}{\xi \bar{\xi} + \eta \bar{\eta}} \right) \frac{2}{\xi' \bar{\xi}' + \eta'\bar{\eta}'} \begin{pmatrix}
\xi' \\ \eta'
\end{pmatrix}  \begin{pmatrix}\bar{\xi}'& \bar{\eta}' \end{pmatrix}
\end{equation}
This gives then, by defining the Möbius transformation $Z_\Lambda := Z^{A_\Lambda}$:
\begin{align}
z_p^\psi(w) &= Z_{\Lambda}(z_p^\psi(v)) \label{tra1}\\
\lambda_p^\psi(w) &=\lambda_p^\psi(v)  \frac{\xi'\bar{\xi}'+ \eta'\bar{\eta}'}{\xi \bar{\xi} + \eta \bar{\eta}} \label{tra2}
\end{align}
Define then for each $A \in \mathrm{PSL}(2, \mathbb{C})$ the map
\begin{equation}
f^A(z) := \frac{\begin{pmatrix} \bar{\xi} &  \bar{\eta}\end{pmatrix}A^* A\begin{pmatrix} \xi \\ \eta \end{pmatrix}}{\begin{pmatrix} \bar{\xi} &  \bar{\eta}\end{pmatrix}\begin{pmatrix} \xi \\ \eta \end{pmatrix}}, \label{RF1}
\end{equation}
which is just the inverse of the conformal factor (\ref{CF}), i.e. $f^A(z) = K^A(z)^{-1}$. We then can write (\ref{tra1}-\ref{tra2}) equally as
\begin{align}
z_p^\psi(w) &= Z_{\Lambda}(z_p^\psi(v)) \label{tra3}\\
\lambda_p^\psi(w) &=\lambda_p^\psi(v) f_\Lambda(z_p^\psi(v))\label{tra4}
\end{align}
where we have defined 
\begin{equation}
f_\Lambda(Z) := f^{A_\Lambda}(z). \label{RF3}
\end{equation}
By this we have derived transformation formulas for coordinate systems $\psi^+\in \mathcal{B}$ under local Lorentz transformations. Now recall, that $\lambda_p^\psi(w) = w^0 = |\vec{w}|$ and $\lambda_p^\psi(v) = v^0 = |\vec{v}|$. By this we then can summarize this finding in a more colloquial language: Any active Lorentz transformation 
\begin{equation}
v^\mu E_\mu \mapsto (\Lambda^\mu_{~\nu} w^\nu) E_\mu 
\end{equation}
corresponds to a unique Möbius transformation
\begin{equation}
z \mapsto Z_\Lambda(z)
\end{equation}
on the Riemann sphere, together with a direction dependent rescaling
\begin{equation}
v^0\mapsto w^0 = v^0 f_\Lambda(\rho(\hat{v})).
\end{equation}
As it stands, it seems a little bit surprising, that the function $f^A$ which determines the rescaling is just the inverse of the conformal factor (\ref{CF}), i.e. $f_A(z) = K_A^{-1}(z)$. But the reason for this was explained in section \ref{EC}.

\section{Prerequisites from group theory}\label{BGT}
We introduce some concepts from group theory, that are needed for our investigation. Especially we will introduce the notion of right semidirect product groups in section \ref{BGT1} and will explain in section \ref{BGT2}, how certain subgroups of right semidirect product groups can be parametrized in terms of so called crossed homomorphisms.
\subsection{Right semidirect product groups}\label{BGT1}
We first introduce the appropriate notion of a semidirect product of groups. Let $P,Q$ be two groups and let
\begin{equation}
\tilde{\kappa}: P \rightarrow \mathrm{Aut}(Q), p \mapsto \tilde{\kappa}_p \label{ah}
\end{equation}
be a group antihomomorphism. The latter is equivalent to the statement, that $\tilde{\kappa}$ defines a right group action of $P$ on $Q$ that acts on $Q$ in terms of automorphisms. We then define the right semidirect product of $P$ and $Q$ with respect to $\tilde{\kappa}$, denoted by $P \ltimes_{\tilde{\kappa}} Q$, as follows (cf. \cite{Col15, sem, sem2}):
\begin{enumerate}
\item The underlying set is given by the cartesian product $P \times Q$.
\item The group operation is given by:
\begin{equation}
(p_1, q_1) (p_2, q_2) = (p_1 p_2, \tilde{\kappa}_{p_2}(q_1) q_2 ) \label{GORS}
\end{equation}
\item The identity element is given by $(e_P, e_Q)$, where $e_P$ and $e_Q$ are the respective identity elements of $P$ and $Q$.
\item The inverse element of $(p,q)$ is given by $(p^{-1}, \tilde{\kappa}_{p^{-1}}(q^{-1}) )$.
\end{enumerate}
Left and right semidirect products are equivalent, in the same way as left and right actions are equivalent (cf. \cite{sem}). Nevertheless, a right semidirect product will be more instructive in the present situation. Please note in addition, that it is of greatest importance, that $\tilde{\kappa}$ acts on $Q$ in terms of automorphisms and especially, that $\tilde{\kappa}_p$ preserves the unit element of $Q$ for any $p \in P$, since otherwise $P \ltimes_{\tilde{\kappa}} Q$ would not even constitute a group. For later use, we also want to define the canonical projection
\begin{equation}
\pi_{\tilde{\kappa}}: P \ltimes_{\tilde{\kappa}} Q \rightarrow P, (p,q) \mapsto p.
\end{equation} 
\subsection{Crossed homomorphisms}\label{BGT2}
Let $P$ denote an arbitrary group, let $Q$ denote an abelian group and let $P \ltimes_{\tilde{\kappa}} Q$ denote their right semidirect product with respect to an antihomomorphism $\tilde{\kappa}: P \rightarrow Q$. Then one can show (cf. \cite{ch, ch2} and p. 88 of \cite{Bro12}), that crossed homomorphisms\footnote{Our definition of a crossed homomorphism is adapted to the occuring right semidirect product. Actually, it should be called a crossed antihomomorphism.}, defined as maps
\begin{equation}
c: P \rightarrow Q, p \mapsto c_p
\end{equation}
which satisfy
\begin{equation}
\forall p_1, p_2 \in P: c_{p_1 p_2} = \tilde{\kappa}_{p_2}(c_{p_1}) c_{p_2},
\end{equation}
parametrize homomorphic sections of the canonical projection
\begin{equation}
\pi_{\tilde{\kappa}}: P \ltimes_{\tilde{\kappa}} Q \rightarrow P.
\end{equation}
This means, they parametrize injective group homomorphisms
\begin{equation}
i_c: P \hookrightarrow P \ltimes_{\tilde{\kappa}} Q
\end{equation}
which satisfy $\pi_{\tilde{\kappa}} \circ i = \mathrm{id}$ (i.e. homomorphic embeddings of the group $P$ into the semidirect product $P \ltimes_{\tilde{\kappa}} Q$) by sending:
\begin{equation}
i_c: P \hookrightarrow P \ltimes_{\tilde{\kappa}} Q, p \mapsto (p, c_p) \label{ch2h}
\end{equation}
By this, there is a $1$-to-$1$-correspondence between subgroups $P \subset P \ltimes_{\tilde{\kappa}} Q$ and crossed homomorphisms $c$ (cf. \cite{ch, Bro12}).
\newpage

\bibliographystyle{plain}
\bibliography{LCSym2}

\end{document}